\documentstyle[12pt]{article}

\begin{document}
\newcommand{\ti}{\theta^{1,0\,\und i}}
\newcommand{\tl}{\theta^{1,0\,\und l}}
\newcommand{\tk}{\theta^{1,0\,\und k}}
\newcommand{\tx}{\theta^{1,0}_{\und i}}
\newcommand{\ta}{\theta^{0,1\,\und a}}
\newcommand{\tb}{\theta^{0,1\,\und b}}
\newcommand{\tc}{\theta^{0,1\,\und c}}
\newcommand{\ty}{\theta^{0,1}_{\und a}}
\newcommand{\da}{D^{2,0}}
\newcommand{\db}{D^{0,2}}
\newcommand{\du}{D^{0,0}_u}
\newcommand{\dv}{D^{0,0}_v}
\newcommand{\qa}{q^{1,0\,\und a}}
\newcommand{\qb}{q^{0,1\,\und i}}
\newcommand{\qc}{q^{\und i\,\und a}}
\newcommand{\nn}{\nonumber}
\newcommand{\be}{\begin{equation}}
\newcommand{\bea}{\begin{eqnarray}}
\newcommand{\eea}{\end{eqnarray}}
\newcommand{\ee}{\end{equation}}
\newcommand{\eps}{\varepsilon}
\newcommand{\und}{\underline}
\newcommand{\p}[1]{(\ref{#1})}
\begin{titlepage}
\begin{flushright}
hep-th/0504185 \\
\end{flushright}
\vskip 1.0truecm
\begin{center}
{\bf ${\cal N}$=8 MECHANICS IN SU(2)xSU(2) HARMONIC SUPERSPACE}
\end{center}
 \vskip 1.0truecm
\centerline{\bf S. Bellucci${}^{\, a,1}$, E. Ivanov${}^{\,b,2}$, A. Sutulin${}^{\,b,2}$}
\vskip 1.0truecm

\centerline{$^a${\it INFN - Laboratori Nazionali di Frascati,}}
\centerline{\it Via E. Fermi 40, P.O. Box
13, I-00044 Frascati, Italy}
\vspace{0.2cm}

\centerline{$^b${\it Bogoliubov Laboratory of
Theoretical Physics, JINR,}}
\centerline{\it 141 980 Dubna, Moscow region,
Russian Federation}
\vskip 1.0truecm  \nopagebreak

\begin{abstract}
\noindent The ${\cal N}{=}8\,, 1D$ analytic bi-harmonic superspace is shown
to provide a natural setting
for ${\cal N}{=}8$ supersymmetric mechanics associated with the off-shell
multiplet ${\bf (4, 8, 4)}$\,. The latter is described by an
analytic superfield $q^{1,1}$\,,
and we construct the general superfield and component actions
for any number of such multiplets.
The set of transformations preserving the flat superspace
constraints on $q^{1,1}$
constitutes ${\cal N}{=}8$ extension of the two-dimensional Heisenberg
algebra {\bf h}(2)\,, with an operator central charge.
The corresponding invariant $q^{1,1}$ action is constructed.
It is unique and breaks $1D$ scale invariance. We also find
a one-parameter family of scale-invariant $q^{1,1}$ actions which,
however, are not invariant under the full ${\cal N}{=}8$ Heisenberg
supergroup. Based on preserving the bi-harmonic Grassmann analyticity,
we formulate ${\cal N}{=}8\,, 1D$ supergravity in terms of
the appropriate analytic
supervielbeins. For its truncated version we construct,
both at the superfield and component levels, the first example
of off-shell $q^{1,1}$
action with local ${\cal N}{=}8\,, 1D$ supersymmetry. This construction can be
generalized to any number of self-interacting $q^{1,1}$\,.

\end{abstract}
\vfill
{\it 1) bellucci@lnf.infn.it}\\
{\it 2) eivanov, sutulin@thsun1.jinr.ru}

\newpage

\end{titlepage}

\section{Introduction}
Supersymmetric quantum mechanics (SQM) \cite{W} and, especially,
its versions with extended $1D$ supersymmetry,
provide a laboratory for exploring characteristic features of the ``parent''
supersymmetric field theories in diverse dimensions (see \cite{Rev} and refs.
therein).
It is also tightly related to string theory and the black holes stuff
as implied by the AdS$_2$/CFT$_1$ correspondence.
SQM models augmented with couplings to the relevant $1D$
supergravities and so possessing local $1D$ supersymmetry amount
to various versions
of the spinning particle models, both in the flat and curved backgrounds
(see e.g.
\cite{{Sor1},{Sor2}})\,.

The study of SQM models with rigid ${\cal N}{=}8\,,
1D$ supersymmetry in various superfield formulations
was initiated in \cite{DE} (see also \cite{{BZ},{Smi1}})
and then continued in \cite{BIKL2} -
\cite{BKS}. In \cite{ABC} the full list of off-shell
${\cal N}{=}8\,, 1D$ multiplets with 8 physical fermions
and finite sets of auxiliary fields was presented. Particular cases
of ${\cal N}{=}8$ SQM associated with some
of these multiplets were a subject of study in recent papers \cite{ISmi} - \cite{BKS}.

The intrinsic geometries of supersymmetric field theories become manifest in the superfield
formulations in which all the underlying supersymmetries are explicit and off-shell.
Here we demonstrate this for ${\cal N}{=}8$ supersymmetric mechanics based on the
${\cal N}{=}8$ multiplet with the off-shell content ${\bf (4, 8, 4)}$ \cite{{ABC},{BKSu}}.
The general action for single such multiplet was recently constructed in \cite{BKSu},
with making use of the Hamiltonian framework and the ${\cal N}{=}4$
superfield formulation in which half
of the underlying supersymmetries are implicit.
One of the purposes of the present paper is to show that
the most appropriate arena for dealing with the ${\bf (4, 8, 4)}$ multiplet
is the ${\cal N}{=}8\,, 1D$
analytic bi-harmonic superspace with two independent sets of the $SU(2)$ harmonic
variables. It is $1D$ reduction of the ${\cal N}{=}(4,4)\,, 2D$ analytic
bi-harmonic superspace \cite{IS,IS1}.
\footnote{The adequacy of ${\cal N}{=}4\,, 1D$ harmonic superspace
with one set of harmonic variables for describing ${\cal N}{=}4$ SQM models
was earlier shown in \cite{IL}.
The standard ${\cal N}{=}8\,, 1D$ HSS obtained by a direct reduction from
the ${\cal N}{=}2\,, 4D$
HSS \cite{{HSS},{HSS1}} was used in \cite{{BZ},{ISmi}} for studying
${\cal N}{=}8$ SQM models
associated with the multiplet ${\bf (5, 8, 3)}\,$.}
The multiplet ${\bf (4,8,4)}$ is described by the analytic bi-harmonic
${\cal N}{=}8$ superfield $q^{1,1}$, which is a reduction of the ${\cal N}{=}(4,4)\,, 2D$
superfield representing one type of twisted ${\cal N}{=}(4,4)$ multiplets.
All eight $1D$ supersymmetries
are manifest in such a formulation. We construct the most general
superfield actions, both for one and few ${\bf (4, 8, 4)}$ multiplets, and
present the corresponding component off-shell and on-shell actions. Besides
the entire ${\cal N}{=}8$ supersymmetry, four $SU(2)$ automorphism
symmetries of the latter are manifest
in the bi-harmonic formulation. Surprisingly, the most general coordinate (and frame)
transformations preserving the flat geometry of ${\cal N}{=}8\,, 1D$
bi-harmonic superspace form ${\cal N}{=}8$ extension of two-dimensional
Heisenberg algebra ${\bf h}(2)$\,, rather than any
of the standard ${\cal N}{=}8$ conformal superalgebras \cite{VPr}.
This should be contrasted with the ${\cal N}{=}(4,4)\,, 2D$ case where
the analogous flat superspace-preserving
algebra is a sum of infinite-dimensional ``large'' ${\cal N}{=}4$ superconformal algebras
of the left and right light-cone $2D$ sectors \cite{IS}.
The corresponding invariant $1D$ action
is constructed and shown to be unique. We also find a family
of dilatation-invariant superfield
actions parametrized by the conformal dimension of $q^{1,1}$\,.
For the generic value of the conformal
dimension these actions inevitably involve non-trivial self-interaction.

As another topic of this paper, we formulate
${\cal N}{=}8$ supergravity (SG) in ${\cal N}{=}8\,, 1D$
analytic harmonic superspace, basically following ref. \cite{BI} where the
analogous setting for the ${\cal N}{=}(4,4)\,, 2D$ SG was developed.
We construct, for the first time, the full off-shell coupling of $q^{1,1}$
to the simplest version of this SG, both in the manifestly supersymmetric
superfield formalism and in the component approach.
Besides local ${\cal N}{=}8$ supersymmetry
and $1D$ diffeomorphisms, this model respects two local $SU(2)$ symmetries
realized on fermions
and two global $SU(2)$ symmetries. An interesting feature of the $1D$ case as compared to
the $2D$ one \cite{BI} is the absence of the einbein field in the original
Weyl-multiplet type
gauge field representations. Two peculiar mechanisms of generating this missing gauge
field are found. One of them can be easily generalized
to the case of generic self-interaction of
any number of $q^{1,1}$ superfields coupled to the ${\cal N}{=}8\,, 1D$ SG.
\setcounter{equation}{0}

\section{$SU(2)\times SU(2)$ harmonic superspace and  multiplet (4,8,4)}

\subsection{Basic definitions}
We begin by defining the
standard real ${\cal N}{=}8$\,, $1D$ superspace that is parametrized
by the following set of coordinates:
$$
{\bf R}^{(1|8)} = (\,Z\,) =
(\,t\,, \theta^{i\, \und k}\,, \theta^{a\, \und b}\,)\,.
$$
This superspace can be also obtained under reduction of ${\cal N}=(4,4)$\,, $2D$
superspace ${\bf R}^{(1,1|4,4)}$\, \cite{IS,IS1}.
Here the indices and $i$\,, $\und k$\,, $a$\,, $\und b$
are doublet indices of four commuting $SU(2)$ groups forming the automorphism
group $SO(4) \times SO(4)$ of the ${\cal N}{=}8, 1D$ superalgebra.
Just this subgroup of the
general ${\cal N}{=}8, 1D$ automorphism group $SO(8)$
is manifest in the considered formulation.
The corresponding covariant spinor derivatives are defined as
\bea
&&
D_{i\,\und k} = \frac{\partial}{\partial \theta^{i\,\und k}} +
i\,\theta_{i\, \und k}\,\partial_t\,, \quad
D_{a\, \und b} = \frac{\partial}{\partial \theta^{a\, \und b}} +
i\,\theta_{a\, \und b}\,\partial_t\,, \nn\\
&&
(D_{i\, \und k})^{\dagger} =
- \eps^{i\,l}\, \eps^{\und k\, \und n}\, D_{l\, \und n}\,, \quad
(D_{a\, \und b})^{\dagger} =
- \eps^{a\,c}\, \eps^{\und b\, \und d}\, D_{c\, \und d}\,,
\eea
and obey the following algebra:
\be
\{\,D_{i\, \und k}\,, D_{j\, \und l} \,\} = 2i\, \eps_{i\,j}\,
\eps_{\und k\, \und l}\,
\partial_t\,, \quad
\{\,D_{a\, \und b}\,, D_{c\, \und d} \,\} = 2i\, \eps_{a\,c}\,
\eps_{\und b\, \und d}\,
\partial_t\,.
\label{alg}
\ee

For the one--dimensional ${\cal N}{=}8$ supersymmetric theory
we can introduce $SU(2)\times SU(2)$ harmonic superspace (HSS) with two independent
sets of harmonic variables $u^{\pm 1}_i$ and $v^{\pm 1}_a$ associated with
two different $SU(2)$ groups of the $SO(4) \times SO(4)$ automorphism group
of the algebra.
This type of HSS is a clear analog of the bi-harmonic superspace introduced in
\cite{IS} to describe off-shell ${\cal N}{=}(4,4)$\,, $2D$ supersymmetric
sigma models
with torsion. As we shall see, it provides the most appropriate framework
for the ${\cal N}{=}8$ supersymmetric quantum mechanics associated
with the $1D$ off-shell supermultiplet ${\bf (4,8,4)}$ \cite{ABC}.

We define the central basis of this HSS as
\be
{\bf HR}^{(1+2+2|8)} = (\,Z\,,u\,,v\,) =
{\bf R}^{(1|8)} \otimes (\,u^{\pm1}_i, v^{\pm1}_a\,)\,, \quad
u^{1i} u^{-1}_i = 1\,,\quad v^{1a} v^{-1} _a = 1\,.
\label{HSS}
\ee
The analytic basis in ${\cal N}{=}8$\,, $SU(2)\times SU(2)$ HSS amounts
to the following choice of coordinates:
\be
{\bf HR}^{(1+2+2|8)} = (\,X\,,u\,,v\,) = (\,t_A\,,
\theta^{\pm 1,0\, \und i}\,, \theta^{0,\pm1\, \und a}\,,
u^{\pm 1}_i\,, v^{\pm 1}_a\,)
\label{an.set}
\ee
where
$$
t_A = t + i (\ti\, \theta^{-1,0}_{\und i} + \ta\, \theta^{0,-1}_{\und a})\,, \quad
\theta^{\pm 1,0\, \und i} = \theta^{k\, \und i}\, u^{\pm 1}_k\,, \quad
\theta^{0,\pm 1\, \und a} = \theta^{b\, \und a}\, v^{\pm 1}_b\,.
$$
The main feature of the analytic basis is
that it visualizes the existence of the {\it analytic subspace} in
the $SU(2)\times SU(2)$ HSS
\be
{\bf AR}^{(1+2+2|4)} = (\,\zeta\,,u\,,v\,)
= (\,t_A\,, \ti\,, \ta\,, u^{\pm 1}_i\,, v^{\pm 1}_a\,)\,,
\label{AS}
\ee
which has twice as less odd coordinates as compared to the standard ${\cal N}{=}8, 1D$
superspace and is closed under ${\cal N}{=}8$ supersymmetry transformations.
The existence of the analytic subspace matches with the form of covariant
spinor derivatives in the analytic basis
\be
D^{1,0\, \und i} = \frac{\partial}{\partial \theta^{-1,0}_{\und i}}\,,\quad
D^{0,1\, \und a} = \frac{\partial}{\partial \theta^{0,-1}_{\und a}}
\label{sp.der}
\ee
where
\be
D^{\pm 1,0\, \und i} \equiv D^{k\, \und i}\, u^{\pm 1}_k\,,\quad
D^{0,\pm 1\, \und a} \equiv D^{b\, \und a}\, v^{\pm 1}_b\,.
\ee
The ``shortness'' of $D^{1,0 \,\und i}\,,
D^{0,1\, \und a}$ means that the Grassmann-analytic bi-harmonic
superfields $\Phi^{\,q,\,p}$,
\be
D^{1,0\, \und i}\, \Phi^{\,q,\,p} = D^{0,1\, \und a}\, \Phi^{\,q,\,p} = 0\,,
\label{Phi}
\ee
do not depend on
$\theta^{-1,0\, \und i}\,, \theta^{0, -1\, \und a}$ in the analytic
basis, i.e., they ``live'' on the analytic superspace \p{AS}:
\be
\Phi^{\,q,\,p} = \Phi^{\,q,\,p} (\zeta, u, v)\,.
\ee
In what follows, for brevity,  we shall frequently omit the index ``A'' on the analytic
basis time coordinate.

In the bi-harmonic superspace one can define two sets of mutually
commuting harmonic derivatives, each forming an $SU(2)$ algebra \cite{IS}.
In the analytic basis, the explicit expressions for the derivatives
with positive $U(1)$ charges, as well as for the derivatives counting
the harmonic $U(1)$ charges $p,q$, when they act on the analytic superfields, read
\bea
&&
\da = \partial^{2,0} + i\, \ti \theta^{1,0}_{\und i} \partial_t\,,\quad
\du = \partial^{0,0}_u + \ti \frac{\partial}{\partial \ti}\,, \nn\\
&&
\db = \partial^{0,2} + i\, \ta \theta^{0,1}_{\und a} \partial_t\,, \quad
\dv = \partial^{0,0}_v + \ta \frac{\partial}{\partial \ta}\,,
\label{harm.der}
\eea
and
\bea
&&
\partial^{2,0} = u^{1i} \frac{\partial}{\partial u^{-1i}}\,, \quad
\partial^{0,0}_u = u^{1i} \frac{\partial}{\partial u^{1i}}
- u^{-1i} \frac{\partial}{\partial u^{-1i}}\,, \nn\\
&&
\partial^{0,2} = v^{1a} \frac{\partial}{\partial v^{-1a}}\,, \quad
\partial^{0,0}_v = v^{1a} \frac{\partial}{\partial v^{1a}} -
v^{-1a} \frac{\partial}{\partial v^{-1a}}\,.
\eea

\subsection{The multiplet (4,8,4)}

In the standard ${\cal N}{=}8, 1D$ superspace ${\bf R}^{(1|8)}$ the multiplet
with the
off-shell field content ${\bf (4,8,4)}$ is described by a real quartet
superfield $q^{\,i\, a}$ subjected to the constraints \cite{ABC}
\be
D^{(k\, \und k} q^{\,i)\, a} = D^{(b\, \und b} q^{\,k\, a)} = 0\,,
\label{q11constr}
\ee
where symmetrization is understood for the doublet indices of the same
automorphism $SU(2)$ group.
On the other hand, one can interpret these constraints as a $1D$ reduction of
the constraints which define ${\cal N}{=}(4,4)$
twisted multiplet in the superspace ${\bf R}^{(1,1|4,4)}$~\cite{IS}.
Since the adequate off-shell description of the twisted multiplet is achieved
in the framework of the bi-harmonic superspace ~\cite{IS},
we wish to use the corresponding techniques in the one--dimensional case
to show that the off-shell multiplet ${\bf (4,8,4)}$ also admits a very simple
description within such framework.

In the superspace ${\bf HR}^{(1+2+2|8)}$  the multiplet ${\bf (4,8,4)}$ can be
described by a real analytic ${\cal N}{=}8$ superfield
$q^{1,1}(\zeta, u, v)$ subjected to the harmonic constraints
\be
\da q^{1,1} = 0\,,\quad \db q^{1,1} = 0\,,
\label{hc}
\ee
which in the central basis imply
\be
q^{1,1} = q^{i\,a}u^1_i v^1_a\,. \nn\\
\ee
Then $q^{ia}$ satisfies the constraints \p{q11constr} as a consequence
of the Grassmann analyticity constraints \p{Phi}.
The analytic basis solution of the harmonic constraints \p{hc} is given by
\bea
q^{1,1} \!&=&\! f^{i\,a}u^1_i v^1_a + \ti \psi^a_{\und i} v^1_a
+ \ta \psi^i_{\und a} u^1_i
- i\, (\theta^{1,0})^2 \partial_t f^{i\,a}u^{-1}_i v^1_a
- i\, (\theta^{0,1})^2 \partial_t f^{i\,a}u^1_i v^{-1}_a \nn\\
\!&+&\!
\ti \ta F_{\und i\, \und a}
- i\, \ti (\theta^{0,1})^2 \partial_t \psi^a_{\und i} v^{-1}_a
- i\, \ta (\theta^{1,0})^2 \partial_t \psi^i_{\und a} u^{-1}_i \nn\\
\!&-&\!
(\theta^{1,0})^2 (\theta^{0,1})^2 \, \partial_t^2 f^{i\,a}u^{-1}_i v^{-1}_a
\label{sol}
\eea
where $(\theta^{1,0})^2 = \theta^{1,0\, \und k}\,\theta^{1,0}_{\und k}$\,,
$(\theta^{0,1})^2 = \theta^{0,1\, \und a}\,\theta^{0,1}_{\und a}\,$.
The independent component $1D$ fields
$f^{ia}, \psi^a_{\und i}, \psi^i_{\und a}, F_{\und i\, \und a}$
form the ${\cal N}{=}8$ off-shell multiplet ${\bf(4, 8, 4)}$.

The general {\it off-shell} action of $n$ such superfields $q^{1,1\, M}$
$(M = 1,2,...n)$ can be written as the following integral over the analytic
superspace (\ref{AS}):
\be
S^{gen} = \int \mu^{-2,-2}\, {\cal L}^{2,2} (q^{1,1\, M},u,v)
\label{s1gen}
\ee
where
\be
\mu^{-2,-2} = dt du\, dv\,d^2\theta^{1,0}\, d^2\theta^{0,1}
\label{measure1}
\ee
is the analytic superspace integration measure normalized as
$$
\int d^2\theta^{1,0}\, d^2\theta^{0,1} (\theta^{1,0})^2 (\theta^{0,1})^2 = 1\,.
$$
The analytic superfield Lagrangian ${\cal L}^{2,2}$ bears in general an arbitrary
dependence on its arguments, the only restriction being a compatibility
with its external $U(1)$ charges $(2,2)$\,.
The free action is given by
\be
S^{free} = \int \mu^{-2,-2}\, q^{1,1\, M} q^{1,1\, M}\,,
\label{free}
\ee
so, for consistency, we are led to assume
$$
\left. \det\left(\frac{\partial^2 {\cal L}^{2,2}}
{\partial q^{1,1\, M}\,\partial q^{1,1\, N}}\right) \right |_{q^{1,1} = 0} \ne 0\,.
$$
Using (\ref{sol})\,, one finds the component form of the action (\ref{free})
\be
S^{free} = \frac{1}{2}\,\int dt\,
\Big \{\, \partial_t f^{i\,a}\,\partial_t f_{i\,a}
+ \frac{i}{2}\,\Big (\,\psi^{\und i\,a}\, \partial_t \psi_{\und i\,a} +
\psi^{i\, \und a}\, \partial_t \psi_{i\, \und a}\, \Big )
+ \frac{1}{4}\, F^{\und i}_{\und a}\, F_{\und i}^{\und a}\, \Big \}\,.
\ee

Passing to the component form of the general action (\ref{s1gen})
is straightforward. We present here both {\it off-shell} and {\it on-shell}
component actions for two cases, when the Lagrangian in (\ref{s1gen}) depends
either on {\it one} $q^{1,1}$ multiplet or on an {\it arbitrary} number
of such multiplets.

In the first case ($n=1$) the off-shell action reads
\bea
S^{gen} \!&=&\! \frac{1}{2} \,\int dt\, \Big \{
G(f)\,\partial_t f^{i\,a}\,\partial_t f_{i\,a}
+ \frac{i}{2}\, G(f)\, \Big(\,\psi^{\und i\,a}\, \partial_t \psi_{\und i\,a} +
\psi^{i\, \und a}\, \partial_t \psi_{i\, \und a}\, \Big ) \nn\\
\!&+&\!
\frac{1}{4}\,G(f)\, F^{\und i}_{\und a}\, F_{\und i}^{\und a}
- \frac{1}{2}\, \frac{\partial G(f)}{\partial f^{i\, a}}\,
\psi^{\und k\, a}\psi^i_{\und b}\, F^{\und b}_{\und k} \nn\\
\!&-&\!
\frac{i}{2}\, \Big (\,\frac{\partial G(f)}{\partial f^{i\, a}}\,
\psi^b_{\und k} \psi^{\und k\, a}\,
\partial_t f^i_b
+ \frac{\partial G(f)}{\partial f^{i\, a}}\,
\psi^k_{\und b} \psi^{i\, \und b}\, \partial_t f^a_k \,\Big )\nn\\
\!&+&\!
\frac{1}{8}\, \frac{\partial^2 G(f)}{\partial f^{i\, a}\, \partial f^{k\, b}}\,
\psi^a_{\und n} \psi^{\und n\, b} \psi^i_{\und d} \psi^{k\, \und d}\,
\Big \}
\label{Scomp1}
\eea
where
\bea
&& G(f) = \int du\,dv\, g(f^{1,1}\,,u\,,v)\,, \quad
g(f^{1,1}\,,u\,,v) = \left. \frac{\partial^2 {\cal L}^{2,2}}{\partial q^{1,1}\,
\partial q^{1,1}}
\right|_{\theta = 0}\,, \nn \\
&& \left. q^{1,1} \right|_{\theta = 0} = f^{1,1} = f^{i\, a}(t) u^1_i v^1_a\,.
\label{g}
\eea
After eliminating the auxiliary fields by their equations of motion
\be
F^{\und k}_{\und b} = G^{-1}(f)\,\frac{\partial G(f)}{\partial f^{i\, a}}\,
\psi^{\und k\, a}
\psi^i_{\und b}\,,
\label{f}
\ee
one obtains the on--shell form of the action (\ref{Scomp1})
\bea
S^{gen} \!&=&\! \frac{1}{2} \, \int dt\, \Big \{
G(f)\,\partial_t f^{i\,a}\,\partial_t f_{i\,a}
+ \frac{i}{2}\, G(f)\, \Big (\, \psi^{\und i\,a}\, \partial_t \psi_{\und i\,a} +
\psi^{i\, \und a}\, \partial_t \psi_{i\, \und a}\, \Big ) \nn\\
\!&-&\!
\frac{i}{2}\, \Big (\, \frac{\partial G(f)}{\partial f^{i\, a}}\,
\psi^b_{\und k} \psi^{\und k\, a}\, \partial_t f^i_b
+ \frac{\partial G(f)}{\partial f^{i\, a}}\, \psi^k_{\und b} \psi^{i\, \und b}\,
\partial_t f^a_k \Big ) \nn\\
\!&+&\!
\frac{1}{8}\, \Big (\,\frac{\partial^2 G(f)}{\partial f^{i\, a}\, \partial f^{k\, b}}\,
 -2\, G^{-1}\, \frac{\partial G(f)}{\partial f^{i\, a}}\,
\frac{\partial G(f)}{\partial f^{k\, b}}\,
\,\Big )
\psi^a_{\und n} \psi^{\und n\, b} \psi^i_{\und d} \psi^{k\, \und d}\,
\Big \}.
\label{onshell1}
\eea
The basic function $G(f)$ in (\ref{g}) satisfies the four--dimensional Laplace equation
\be
\bigtriangleup G(f) = 0\,, \quad
\bigtriangleup = \frac{\partial^2}{\partial f^{i\,a}\, \partial f_{i\,a}}\,
\label{Lap}
\ee
which follows from the definition of $G(q)$\,. This general component action coincides
with that obtained in \cite{BKS}, within the ${\cal N}{=}4, 1D$ superfield formulation with
only four out of eight supersymmetries being manifest, by thoroughly
studying the restrictions imposed
by four hidden supersymmetries. We see that the manifestly ${\cal N}{=}8$ supersymmetric
approach immediately yields this action.

In the case when the Lagrangian depends on a few $q^{1,1\,M}$ multiplets, ($M=1,...,n$),
the component off-shell action reads
\bea
S^{Gen} \!&=&\! \frac{1}{2} \,\int dt\, \Big \{
G_{M\,N}(f)\,\partial_t f^{i\,a\,M}\,\partial_t f_{i\,a}^N
+ \frac{i}{2}\, G_{M\,N}(f)\, \Big (\,\psi^{\und i\,a\,M}\, \partial_t \psi_{\und i\,a}^N +
\psi^{i\, \und a\,M}\, \partial_t \psi_{i\, \und a}^N\, \Big) \nn\\
\!&+&\! \frac{1}{4}\,G_{M\,N}(f)\, F^{\und i\,M}_{\und a}\, F_{\und i}^{\und a\,N}
- \frac{1}{2}\, \frac{\partial G_{M\,N}(f)}{\partial f^{i\, a\,T}}\,
\psi^{\und k\, a\,M} \psi^{i\,N}_{\und b}\, F^{\und b\,T}_{\und k} \nn\\
\!&-&\! \frac{i}{2}\,
\Big (\, \frac{\partial G_{M\,N}(f)}{\partial f^{i\, a\,T}}\,
\psi^{b\,N}_{\und k} \psi^{\und k\, a\,T}\, \partial_t f^{i\,M}_b
+ \frac{\partial G_{M\,N}(f)}{\partial f^{i\, a\,T}}\,
\psi^{k\,N}_{\und b} \psi^{i\, \und b\,T}\, \partial_t f^{a\,M}_k \,\Big ) \nn\\
\!&+&\!
\frac{1}{8}\, \frac{\partial^2 G_{M\,N}(f)}{\partial f^{i\, a\,T}\, \partial f^{k\, b\,L}}\,
\psi^{a\,M}_{\und n} \psi^{\und n\, b\,T} \psi^{i\,N}_{\und d} \psi^{k\, \und d\,L}\,
\Big \}
\label{Scomp2}
\eea
where
\bea
G_{M\,N}(f) \!&=&\! \int du\, dv\, g_{M\,N}(f^{1,1\,M}\,,u\,,v)\,, \quad
g_{M\,N}(f^{1,1\,M}\,,u\,,v) =
\left. \frac{\partial^2 {\cal L}^{2,2}}{\partial q^{1,1\,M}\, \partial q^{1,1\,N}}
\right|_{\theta = 0}\,, \nn\\
\left. q^{1,1\,M} \right|_{\theta = 0}
\!&=&\! f^{1,1\,M} \!=\! f^{i\, a\, M}(t) u^1_i v^1_a\,.
\label{G}
\eea
Then, eliminating the auxiliary fields $F^{\und i\,M}_{\und a}$ by their equations of motion
\be
F^{\und k\, L}_{\und b} = G^{L\,T}(f)\,
\frac{\partial G_{M\,N}(f)}{\partial f^{i\, a\, T}}\, \psi^{\und k\, a\,M}
\psi^{i\,N}_{\und b}\,, \quad G^{L\,T}(f)\, G_{T\,M}(f) = \delta^L_M\,,
\ee
(here $G^{L\,T}(f)$ is an inverse metric -- the analog of $G^{-1}$ in (\ref{f}))
one finds the on-shell form of the action (\ref {Scomp2})
\bea
S^{Gen} \!&=&\! \frac{1}{2} \,\int dt\, \Big \{
G_{M\,N}(f)\,\partial_t f^{i\,a\,M}\,\partial_t f_{i\,a}^N
+ \frac{i}{2}\, G_{M\,N}(f)\,( \psi^{\und i\,a\,M}\, \partial_t \psi_{\und i\,a}^N +
\psi^{i\, \und a\,M}\, \partial_t \psi_{i\, \und a}^N ) \nn\\
\!&-&\!
\frac{i}{2}\,
\Big (\, \frac{\partial G_{M\,N}(f)}{\partial f^{i\, a\,T}}\,
\psi^{b\,N}_{\und k} \psi^{\und k\, a\,T}\, \partial_t f^{i\,M}_b
+ \frac{\partial G_{M\,N}(f)}{\partial f^{i\, a\,T}}\,
\psi^{k\,N}_{\und b} \psi^{i\, \und b\,T}\, \partial_t f^{a\,M}_k \, \Big ) \nn\\
\!&+&\!
\frac{1}{8}\,\Big (\,\frac{\partial^2 G_{M\,N}(f)}{\partial f^{i\, a\,T}\,
\partial f^{k\, b\,L}}\,
-2\,G^{I\,J}(f)\,
\frac{\partial G_{M\,N}(f)}{\partial f^{i\, a\,I}}\,
\frac{\partial G_{T\,L}(f)}{\partial f^{k\, b\,J}}\,
\,\Big )
\psi^{a\,M}_{\und n} \psi^{\und n\, b\,T} \psi^{i\,N}_{\und d} \psi^{k\, \und d\,L}\,
\Big \}.
\label{onshell2}
\eea

The analog of the scalar function $G(f)$ of the one-multiplet
case is the symmetric $n\times n$ matrix function $G_{M\,N}(f)$.
{}From its definition (\ref{G}) it is easy
to find analogs of the constraint (\ref{Lap}) for the considered case
\be
(\mbox{a}) \;\;\; \frac{\partial^2 G_{M\,N}(f)}{\partial f^{i\, a \,T}\,
\partial f_{i\,a}^L} = 0\,,
\quad (\mbox{b}) \;\;\;\frac{\partial G_{M\,N}(f)}{\partial f^{i\, a\,T}} -
\frac{\partial G_{T\,N}(f)}{\partial f^{i\, a\,M}}\,= 0\,.
\label{constrMN}
\ee
Thus, in the present case we are facing the same type of bosonic target
HKT (hyper-K\"ahler with torsion) geometry as in the $2D$ case \cite{HKT}.

\setcounter{equation}{0}

\section{``Superconformal'' group of $SU(2)\times SU(2)$ $1D$ HSS}
\subsection{The supergroup preserving flat harmonic derivatives}

In ${\cal N}{=}(4,4)$, $2D$ bi-harmonic superspace the requirement of preserving the flat
form of the harmonic derivatives $D^{2,0}$, $D^{0,2}$ uniquely selects
the infinite-dimensional ``large'' ${\cal N}{=}4$ superconformal groups
(both in the left and right
light cone sectors) as the most general coordinate groups meeting this requirement.
Surprisingly,
the same requirement in the $1D$ version of bi-harmonic superspace, together
with the demand of covariance of the defining $q^{1,1}$ constraints \p{hc},
pick up a supergroup which does
not coincide with any known $1D$ superconformal group \cite{VPr}.
Instead, it is a ${\cal N}{=}8$
superextended Heisenberg group with an operator central charge.

The 1D version of the general 2D superdiffeomorphism group which
preserves the bi-harmonic analyticity and the defining conditions of the harmonics
$u^{1 i}u^{-1}_i = 1\,, v^{1 a}v^{-1}_a = 1\;$ can be shown to act on the coordinates
of the superspace ${\bf AR}^{(1+2+2|4)}$ as
\bea
&&
\delta u^1_i = \Lambda^{2,0} u^{-1}_i\,,\quad
\delta u^{-1}_i = 0\,,\quad
\delta v^1_a = \Lambda^{0,2} v^{-1}_a\,,\quad
\delta v^{-1}_a = 0\,, \nn \\
&&
\delta \ti = \Lambda^{1,0\, \und i}\,, \quad
\delta \ta = \Lambda^{0,1\, \und a}\,, \quad
\delta t = \tilde \Lambda\,.
\label{tran1}
\eea
The requirement that the harmonic derivatives $\da$\,, $\db$ preserve
their form under these transformations makes them transform as
\be
\delta \da = -\Lambda^{2,0} \du\,, \quad
\delta \db = -\Lambda^{0,2} \dv\,,
\ee
and leads to the following constraints on the parameters $\Lambda$:
\bea
&&
\da \Lambda^{2,0} = 0\,,\quad
\da \Lambda^{0,2} = 0\,,\nn\\
&&
\da \Lambda^{1,0\, \und i} = \ti \Lambda^{2,0}\,, \quad
\da \Lambda^{0,1\, \und a} = 0\,,\quad
\da \tilde \Lambda = - 2i\, \theta^{1,0}_{\und i}
\Lambda^{1,0\, \und i}\,, \nn\\
&&
\db \Lambda^{2,0} = 0\,,\quad
\db \Lambda^{0,2} = 0\,,\nn\\
&&
\db \Lambda^{1,0\, \und i} = 0\,,\quad
\db \Lambda^{0,1\, \und a} = \ta \Lambda^{0,2}\,, \quad
\db \tilde \Lambda = - 2i\, \theta^{0,1}_{\und a} \Lambda^{0,1\, \und a}\,.
\label{con.lambda}
\eea
The general solution to eqs. (\ref{con.lambda})
is provided by
\bea
&&
\Lambda^{2,0} = \lambda^{(ij)} u^1_i u^1_j + \ti \lambda^k_{\und i} u^1_k
+ (\theta^{1,0})^2 \lambda_0\,, \nn\\
&&
\Lambda^{0,2} = \lambda^{(ab)} v^1_a v^1_b + \ta \lambda^b_{\und a} v^1_b
+ (\theta^{0,1})^2 \tilde \lambda_0\,, \nn\\
&&
\Lambda^{1,0\, \und i} = \eps^{k\, \und i} u^1_k
+ \theta^{1,0\, \und k}\, (\delta^{\und i}_{\und k}\,\alpha_0
+ \lambda^{\und i\,)}_{(\, \und k}
+ \delta^{\und i}_{\und k}\,\lambda^{(ij)} u^1_i u^{-1}_j)
- \frac{1}{2} (\theta^{1,0})^2 \lambda^{k\,\und i} u^{-1}_k\,, \nn\\
&&
\Lambda^{0,1\, \und a} = \eps^{b\, \und a} v^1_b
+ \theta^{0,1\, \und b}\, (\delta^{\und a}_{\und b}\,\beta_0
+ \lambda^{\und a\,)}_{(\, \und b}
+ \delta^{\und a}_{\und b}\, \lambda^{(ab)} v^1_a v^{-1}_b)
- \frac{1}{2} (\theta^{0,1})^2 \lambda^{b\,\und a} v^{-1}_b\,, \nn\\
&&
\tilde \Lambda = p_0(t) + 2i\, \ti\, \eps^k_{\und i} u^{-1}_k
+ 2i \, \ta\, \eps^b_{\und a} v^{-1}_b
+ i\, (\theta^{1,0})^2 \lambda^{(ij)} u^{-1}_i u^{-1}_j
+ i\, (\theta^{0,1})^2 \lambda^{(ab)} v^{-1}_a v^{-1}_b\,, \nn\\
&&
\alpha_0 = \beta_0 = \frac{1}{2}\, \partial_t\, p_0 (t)\,, \quad
p_0 (t) = p_0 + 2 \alpha_0 t
\label{param}
\eea
where all parameters are independent of $t$\,.\footnote{Actually, $D^{2,0}$ and $D^{0,2}$
retain their flat form also under the gauge transformations
$\delta u^{-1}_i = \Lambda^{-2,0}u^{1}_i\,,\delta v^{-1}_a = \Lambda^{0,-2}v^{1}_a\,,
\delta u^{1}_i = \delta v^{1}_a =0\,, \delta t_A =i(\theta^{1,0})^2\Lambda^{-2,0}
+ i(\theta^{0,1})^2\Lambda^{0,-2}$ where $\Lambda^{-2,0}, \Lambda^{0,-2}$ are
unconstrained analytic functions.
This group is also compatible with the defining relations for harmonics.
It mixes $D^{2,0}$ and $D^{0,2}$ among themselves and so
preserves the $q^{1,1}$ constraints \p{hc}, provided that $q^{1,1}$
behaves as a scalar. However, the component fields in $q^{1,1}$
are fully inert under this group, so it bears no interest for our consideration.
It is analogous to the group of arbitrary harmonic $U(1)$ transformations.}

Let us discuss the transformation properties of $q^{1,1}$ under
the ``would-be superconformal'' group defined in
(\ref{tran1})\,, (\ref{param})\,. This transformation law is uniquely fixed by the requirement
that the harmonic constraints (\ref{hc}) are covariant with respect to it.

The variation of $q^{1,1}$ under the ``superconformal'' transformations
can be taken in the form
\be
\delta q^{1,1} = \Lambda\, q^{1,1}\,,\label{TRANq}
\ee
where $\Lambda$ is, for the time being, an arbitrary analytic superfunction.
Then, the conditions of covariance of the constraints (\ref{hc}),
\be
\delta (\da q^{1,1})\; \sim \;\da q^{1,1}\,, \quad
\delta (\db q^{1,1})\; \sim \; \;\db q^{1,1}\,,
\ee
impose the relations
\bea
\Lambda^{2,0} = \da \Lambda\,, \quad \Lambda^{0,2} = \db \Lambda\,, \label{LL}
\eea
whence $\Lambda$ is fixed up to an arbitrary constant $\beta$
\bea
\Lambda = \beta - i\, t\,\lambda_0 + \lambda^{(ij)} u^1_i u^{-1}_j
+ \lambda^{(ab)} v^1_a v^{-1}_b + \ti \lambda^k_{\und i} u^{-1}_k
+ \ta \lambda^b_{\und a} v^{-1}_b\,. \label{LAMBD}
\eea
Note that the relation \p{LL} implies $\tilde\lambda_0 = \lambda_0$ in $\Lambda^{2,0}$
and $\Lambda^{0,2}$ in \p{param}.  The meaning of the new parameter $\beta$ can be clarified
by evaluating the Lee bracket between the transformations of
ordinary ${\cal N}{=}4$ supersymmetry (with parameters
$\eps^{k\, \und i} , \eps^{a\, \und b}$)
and those of the additional ``special'' supersymmetry
(with parameters $\lambda^{k\,\und i}, \lambda^{a\,\und b}$),
as applied to the superfield $q^{1,1}$. The closure of these transformations contains,
besides four mutually commuting $SU(2)$ transformations, also a rescaling of $q^{1,1}$
by a constant which can be identified with $\beta$ in \p{LAMBD}.
This additional parameter never appears in any Lie bracket of the coordinate transformations,
and the related transformation commutes with all other ones.
So the generator of this transformation
can be regarded as an operator central charge $Z$ and can be normalized so that
\be
Z q^{1,1} = q^{1,1}\,.
\ee
One can also ascribe to $q^{1,1}$ an arbitrary weight $k$ under dilatations with the
parameter $\alpha_0$, which corresponds to the following additional shift of $\Lambda$:
\be
\Lambda' = \Lambda + k\alpha_0 \label{LAMBDprime}
\ee
A difference from the central charge $Z$ is that the generator of dilatations has
a nontrivial realization on the superspace coordinates as well.
Note that the analytic superspace integration measure is invariant under
all the above transformations, except for the dilatations:
\be
\delta \mu^{-2,-2} = -2 \alpha_0\, \mu^{-2,-2}\,.
\label{measure}
\ee

The superalgebra behind the above transformations can be revealed by representing the
``active'' variation of $q^{1,1}$ as
\be
\delta^* q^{1,1} = -\gamma^A F_A^M(z)\partial_M q^{1,1} + \gamma^A F_A(z) q^{1,1}
\equiv -\gamma^A Q_A q^{1,1}\,,
\ee
where
\be
\delta z^M \equiv \gamma^A F_A^M(z)\,, \;\; \Lambda' \equiv \gamma^A F_A(z)\,, \;\;
z^M \equiv (\,t_A\,, \ti\,, \ta\,, u^{\pm 1}_i\,, v^{\pm 1}_a\,)\,,
\ee
and $\gamma^A$ and $Q_A$ stand for the infinitesimal parameters and relevant generators.

The explicit form of the generators with the non-zero weight pieces is as follows:
\bea
&&
\tilde D = D - k \,, \quad
\tilde K = K +i t Z\,, \nn\\
&&
\tilde A_{(k\,l)} = A_{(k\,l)} - u^1_{(k} u^{-1}_{l)} Z\,, \quad
\tilde B_{(a\,b)} = B_{(a\,b)} - v^1_{(a} v^{-1}_{b)} Z\,, \nn\\
&&
\tilde S_{k\, \und i} = S_{k\, \und i} - u^{-1}_k \theta^{1,0}_{\und i} Z\,, \quad
\tilde S_{b\, \und a} = S_{b\, \und a} - v^{-1}_b \theta^{0,1}_{\und a} Z\,.
\eea
The differential parts of these generators, as well as the generators possessing
no weight pieces at all, are given by
\bea
&&
P = \partial_t\,, \qquad
D = 2t\, \partial_t
+ \ti\, \frac{\partial}{\partial \ti} + \ta\, \frac{\partial}{\partial \ta}\,, \nn\\
&&
K = (\theta^{1,0})^2\, \partial^{-2,0} + (\theta^{0,1})^2\, \partial^{0,-2}\,, \nn\\
&&
Q_{k\, \und i} = u^1_k\,\frac{\partial}{\partial \ti}
+ 2i\, u^{-1}_k\, \theta^{1,0}_{\und i} \partial_t\,, \nn\\
&&
Q_{b\, \und a} = v^1_b\, \frac{\partial}{\partial \ta}
+ 2i\, v^{-1}_b\, \theta^{0,1}_{\und a} \partial_t\,, \nn\\
&&
S_{k\, \und i} = u^1_k\, \theta^{1,0}_{\und i}\, \partial^{-2,0}
- \frac{1}{2}\, u^{-1}_ k\, (\theta^{1,0})^2\, \frac{\partial}{\partial \ti}\,, \nn\\
&&
S_{b\, \und a} = v^1_b\, \theta^{0,1}_{\und a}\, \partial^{0,-2}
- \frac{1}{2}\, v^{-1}_b\, (\theta^{0,1})^2\, \frac{\partial}{\partial \ta}\,, \nn\\
&&
A_{(k\,l)} = u^1_k u^1_l \partial^{-2,0} + u^1_{(k} u^{-1}_{l)}\,
\ti\,\frac{\partial}{\partial \ti}
+ i\,u^{-1}_k u^{-1}_l\,(\theta^{1,0})^2\,\partial_t\,, \nn\\
&&
A_{(\und k\, \und l)} = - \frac{1}{2}\, \Big [\,
\theta^{1,0}_{\und k}\,\frac{\partial}{\partial \tl}
+ \theta^{1,0}_{\und l}\,\frac{\partial}{\partial \tk}\, \Big]\,, \nn\\
&&
B_{(a\,b)} = v^1_a v^1_b \partial^{0,-2} +  v^1_{(a} v^{-1}_{b)}\,
\ta\, \frac{\partial}{\partial \ta}
+i\, v^{-1}_a v^{-1}_b\, (\theta^{0,1})^2\, \partial_t\,, \nn\\
&&
B_{(\und a\, \und b)} = - \frac{1}{2}\, \Big [\,
\theta^{0,1}_{\und a}\,\frac{\partial}{\partial \tb}
+ \theta^{0,1}_{\und b}\,\frac{\partial}{\partial \ta}\,
\, \Big]\,.
\eea

The non-vanishing (anti)commutators of the closed superalgebra constituted
by these generators read
\bea
&&
[\,P, \tilde D\,] = 2 P\,, \quad [\,\tilde K, \tilde D\,] = -2 \tilde K\,, \quad
[\,P, \tilde K\,] = i\, Z\,,\nn\\
&&
[\,Q_{k\, \und i}\,, \tilde D\,] = Q_{k\, \und i}\,, \quad
[\,Q_{b\, \und a}\,,\tilde D\,] = Q_{b\, \und a}\,, \quad
[\,\tilde S_{k\, \und i}\,, \tilde D\,] = - \tilde S_{k\, \und i}\,, \quad
[\,\tilde S_{b\, \und a}\,,\tilde D\,] = - \tilde S_{b\, \und a}\,, \nn\\
&&
[\,Q_{k\, \und i}\,,\tilde K\,] = 2\,\tilde S_{k\, \und i}\,, \quad
[\,Q_{b\, \und a}\,,\tilde K\,] = 2\,\tilde S_{b\, \und a}\,,\nn\\
&&
[\,Q_{k\, \und i}\,,\tilde A_{(l\,n)}\,]
= \frac{1}{2}\,\Big (\,
\eps_{k\,n}\,Q_{l\, \und i} +  \eps_{k\,l}\,Q_{n\, \und i}
\,\Big )\,,
\nn\\
&&
[\,Q_{b\, \und a}\,, \tilde B_{(c\,d)}\,] = \frac{1}{2}\,\Big (\,
\eps_{b\,d}\,Q_{c\, \und a} + \eps_{b\,c}\,Q_{d\, \und a}
\,\Big )\,, \nn\\
&&
[\,\tilde S_{k\, \und i}\,,\tilde A_{(l\,n)}\,]
= \frac{1}{2}\,\Big (\,
\eps_{k\,n}\,\tilde S_{l\, \und i} +  \eps_{k\,l}\,\tilde S_{n\, \und i}
\,\Big )\,,\quad
[\,\tilde S_{k\, \und i}\,,A_{(\und l\, \und n)}\,] = \frac{1}{2}\,\Big (\,
\eps_{\und i\, \und l}\,\tilde S_{k\, \und n} +  \eps_{\und i\, \und n}\,\tilde S_{k\, \und l}
\,\Big )\,,
\nn\\
&&
[\,\tilde S_{b\, \und a}\,,\tilde B_{(c\,d)}\,] = \frac{1}{2}\,\Big (\,
\eps_{b\,d}\,\tilde S_{c\, \und a} + \eps_{b\,c}\,\tilde S_{d\, \und a}
\,\Big )\,, \quad
[\,\tilde S_{b\, \und a}\,,B_{(\und c\, \und d)}\,] = \frac{1}{2}\,\Big (\,
\eps_{\und a\, \und c}\,\tilde S_{b\, \und d} +  \eps_{\und a\, \und d}\,\tilde S_{b\, \und c}
\,\Big )\,,
\nn\\
&&
\{\,Q_{k\, \und i}\,,\tilde S_{l\, \und n} \,\} =
\eps_{k\,l}\, A_{(\und n\, \und i)}
+ \eps_{\und n\, \und i}\, \tilde A_{(l\,k)}
+ \frac{1}{2}\, \eps_{\und n\, \und i}\, \eps_{l\,k}\,Z\,, \nn\\
&&
\{\,Q_{b\, \und a}\,,\tilde S_{c\, \und d}\, \} =
\eps_{b\,c}\, B_{(\und d\, \und a)}
+ \eps_{\und d\, \und a}\, \tilde B_{(c\,b)}
+ \frac{1}{2}\, \eps_{\und d\, \und a}\, \eps_{c\,b}\,Z\,, \nn \\
&&
\{\,Q_{k\, \und i}\,,Q_{l\, \und n} \,\} = - 2i\, \eps_{k\,l}\,
\eps_{\und i\, \und n}\,P\,, \quad
\{\,Q_{b\, \und a}\,,Q_{c\, \und d}\, \} = -2i\, \eps_{b\,c}\, \eps_{\und a\, \und d}\,P\,, \nn\\
&&
[\,Q_{k\, \und i}\,,A_{(\und l\, \und n)}\,] = \frac{1}{2}\,\Big (\,
\eps_{\und i\, \und l}\,Q_{k\, \und n} +  \eps_{\und i\, \und n}\,Q_{k\, \und l}
\,\Big )\,, \nn\\
&&
[\,Q_{b\, \und a}\,,B_{(\und c\, \und d)}\,] = \frac{1}{2}\,\Big (\,
\eps_{\und a\, \und c}\,Q_{b\, \und d} +  \eps_{\und a\, \und d}\,Q_{b\, \und c}
\,\Big )\,.
\label{algcon1}
\eea

Inspecting the dimensions of the involved generators and their (anti)commutators
we observe that \p{algcon1} in some respects resembles some
${\cal N}{=}8, 1D$ superconformal algebra. Indeed, the generators $K$
and $\tilde{S}$ are analogs
of the generators of $1D$ conformal boosts and conformal
supersymmetry, four R-symmetry $SU(2)$
algebras appear in the anticommutators of the
${\cal N}=8, 1D$ Poincar\'e supersymmetry generators
$Q_{k\, \und i}, Q_{a\, \und b}$ with those of ``conformal''
supersymmetry  $\tilde{S}_{k\, \und i},
\tilde{S}_{a\, \und b}$, etc. The crucial difference is, however,
that the generators $P$ and $\tilde{K}$ together
with the central charge $Z$ form a two-dimensional Heisenberg algebra ${\bf h}(2)$
(or ``the magnetic translations'' algebra \cite{Japan}) rather than the $1D$ conformal
algebra $sl(2,R) \sim so(1,2)$. Besides, the ``conformal'' supersymmetry generators
anticommute with each other, the generator of
dilatations decouples from the remaining generators (it forms an ideal) and, finally,
the central charge $Z$ appears in the $\{Q, \tilde{S} \}$ anticommutators. So, \p{algcon1}
should be treated as a ${\cal N}{=}8$ superextension of the simplest (two-generator)
Heisenberg algebra, rather than any type of superconformal algebra.
It is an open question, whether it can be recovered as a contraction
of any known ${\cal N}{=}8, 1D$ superconformal algebras \cite{VPr}, despite
the fact that the Heisenberg
algebra on its own can be treated as some contraction of $sl(2,R)$ \cite{{Japan},{ASo}}.

To avoid a misunderstanding, let us point out that general diffeomorphisms of
the analytic superspace $(\zeta, u, v)$ still contain as subgroups two
infinite-dimensional ``large'' ${\cal N}{=}4$ superconformal groups which act on the
coordinate subsets $(\,t_A\,, \ti\,, u^{\pm 1}_i\,)$ and $ (\,t_A\,, \ta\,, v^{\pm 1}_a\,)$
and transformations of which have the same form as in the ${\cal N}{=}(4,4), 2D$ case
\cite{{IS},{IS1},{BI}}. However, each of these groups preserves the flat form
of only one of two
harmonic derivatives $D^{2,0}$, $D^{0,2}$,
but not the flat form of two derivatives simultaneously,
as in the ${\cal N}{=}(4,4), 2D$ case. The reason is that in the $1D$ case
both derivatives contain the partial derivative with respect to the same $t_A$, while in the
$2D$ case - with respect to two independent $2D$ light-cone coordinates $x^{++}$ or $x^{--}$.
The maximal subgroup simultaneously preserving {\it both} flat harmonic derivatives of
the $1D$ bi-harmonic superspace is defined by the variations \p{param},
and the algebra of these transformations is finite-dimensional and
is given by the relations \p{algcon1}.

The last comment concerns the relation to the possible interpretation
of the multiplet $({\bf 4, 8, 4})$, along the lines of ref.~\cite{IKL2},
as the Goldstone multiplet parametrizing the appropriate coset
of the ${\cal N}{=}8, 1D$ superconformal group $OSp(4^*|4)$ \cite{ABC}.
This supergroup definitely admits a realization on the coordinates $Z$ of the
standard ${\cal N}{=}8, 1D$ superspace ${\bf R}^{(1|8)}$ and on the constrained superfield
$q^{ia}(Z)$ representing the multiplet $({\bf 4,8,4})$ in ${\bf R}^{(1|8)}$
(see \p{q11constr}). So there arises the question as to why this superconformal group
does not show up
in the analytic superspace description of the multiplet $({\bf 4, 8, 4})$, i.e. why
it is absent in the set of coordinate transformations preserving the flat form of
$D^{2,0}, D^{0,2}$. The reason is that passing to the bi-harmonic extension of
${\bf R}^{(1|8)}$ reduces the general R-symmetry group  $SO(8)$ of ${\bf R}^{(1|8)}$
down to its subgroup $SO(4)\times SO(4)$, while no ${\cal N}{=}8, 1D$ superconformal
groups with such R-symmetry exist \cite{VPr}. In particular, R-symmetry subgroup of
$OSp(4^*|4)$ is $USp(4)\times SU(2) \sim SO(5)\times SU(2)$. Hence,
in the 1D bi-harmonic superspace it is impossible to realize any standard
${\cal N}{=}8$ superconformal group, under the assumption that the corresponding
R-symmetry group acts {\it linearly} on the harmonic variables
$u^{\pm 1}_i, v^{\pm 1}_a$. On the other hand, having Goldstone
$({\bf 4,8, 4})$ multiplet, with physical bosons parametrizing the R-symmetry
coset $SO(5)/SO(4)$, one can realize the R-symmetry $SO(5)$ on the harmonic variables
by the transformations which are {\it nonlinear} in these physical bosonic
fields. Hopefully, this extends to the whole $OSp(4^*|4)$ group which thus could admit
a realization in the analytic bi-harmonic superspace, such that the corresponding
coordinate variations involve the superfield $q^{1,1}$ itself. Such nonlinear realizations
are beyond the scope of our consideration here and will be studied elsewhere.\footnote{E.I.
thanks S. Krivonos for useful correspondence on these points.}

\subsection{Invariant actions}
Here we construct the superfield actions invariant under the group
\p{tran1}, \p{param}, \p{TRANq},
\p{LAMBD}, \p{LAMBDprime}. We follow the method of ref. \cite{IS}.

Let us introduce the superfield
\be
\tilde q^{1,1} = q^{1,1} - c^{1,1}
\ee
where $c^{1,1} = c^{ia}u^1_iv^1_a$ and $c^{ia}$ is a quartet of constants:
\be
c^{1,1}\,c^{-1,-1} - c^{1,-1}\,c^{-1,1} = \frac{1}{2}\, c^2\,, \quad
c^2 = c^{ia}c_{ia}\neq 0\,.
\ee

This newly defined quantity has an inhomogeneous transformation law
under the action of the supergroup considered in the previous Subsection
\be
\delta \tilde q^{1,1} = \Lambda'\, (\tilde q^{1,1} + c^{1,1})
- \Lambda^{2,0}\, c^{-1,1} - \Lambda^{0,2}\, c^{1,-1}\,.
\ee
Let us recall that $\Lambda' = \Lambda + k\,\alpha_0$. We firstly consider
the transformations with the superparameter $\Lambda$ and, separately,
the dilatations with the parameter $\alpha_0$.

The superspace action invariant under the $\Lambda$ transformations
can be sought as a series in $\tilde q^{1,1}$
\be
S_h = \int \mu^{-2,-2}\, \sum_{n=2}^{\infty}\, b_n\,
(\tilde q^{1,1})^n\, (c^{-1,-1}))^{n-2}\,.
\label{ac}
\ee
The integration measure is invariant under the transformations
with the parameters
collected in $\Lambda$\,.
Calculating the variation of the action (\ref{ac}) under $\Lambda$\,,
one finds the following recurrence relations between the coefficients $b_n$:
\be
b_{n+1} = - \frac{2}{c^2}\, \frac{n^2}{(n+1)(n-1)}\, b_n \;\;\;
\Rightarrow\;\;\;
b_n = \Big (- \frac{2}{c^2}\, \Big )^{n-2}\,\frac{2(n-1)}{n}\, b_2\,.
\label{rec}
\ee
Without loss of generality, in what follows we put $b_2 = \frac 12\,$.
Introducing
\be
y = \frac{2}{c^2}\,\tilde q^{1,1}\, c^{-1,-1}\,,
\ee
it is straightforward to show that the series in (\ref{ac}) is summed
up into the expression
\be
S_{h} = \int \mu^{-2,-2}\,\tilde q^{1,1}\,\tilde q^{1,1}\, R(y)\,
\ee
where
\be
R(y) = \frac{\ln(1+y)}{y^2} - \frac{1}{y(1+y)}\,. \label{WZW}
\ee
This superfield Lagrangian formally coincides with the Lagrangian of the
${\cal N}{=}(4,4), 2D$ Wess-Zumino-Witten model found in \cite{IS}, though
it lacks $1D$ scale and conformal invariance. The corresponding metric function $G_h(f)$
specifying the component action (see \p{Scomp1}, \p{g}) is given by
\be
G_h(f) = \int du dv\, \frac{1}{(1 + \hat{y})^2} = \frac{c^2}{\tilde{f}^2
+ 2(\tilde{f} \cdot c) +
c^2} = \frac{c^2}{f^2}\,, \quad \hat{y} \equiv y|_{\theta=0}\,
\label{100}
\ee
where $\tilde f^2 = \tilde f^{i\,a} \tilde f_{i\,a}$\,,
$(\tilde f \cdot c)= \tilde f^{i\,a} c_{i\,a}$\,.
The invariance under the $SU(2)\times SU(2)$ group acting on the indices $i, a$ is
obvious from this representation.

Next, we are going to find an action which is invariant under scale transformations
with the parameter $\alpha_0$\,. In this case the analytic measure
is transformed as in (\ref{measure}) and the recurrence relations between
the coefficients $b_n$ in (\ref{ac}) look like
\be
b_{n+1} = - \frac{2}{c^2}\, \frac{n(n-\rho)}{(n+1)(n-1)}\, b_n\,, \quad
\rho = \frac{2}{k}\,.
\label{gamma}
\ee
Note that now the recurrence relations involve an arbitrary constant
$\rho$\,. The formulas (\ref{rec})\,, (\ref{gamma}) coincide only for a singular
choice
$$
\rho = 0\; (k = \infty)\,.
$$

{}From (\ref{gamma}) one finds
\be
b_n = \Big (-\frac{2}{c^2}\, \Big)^n \frac{2(n-1-\rho)!}{n(n-2)!\,(1-\rho)!}\, b_2\,,
\quad n\geq 2\,.
\ee
Then, the one-parameter family of actions invariant under dilatations takes the form
\be
S_{scale}(\rho) = \int \mu^{-2,-2}\, \tilde q^{1,1}\,\tilde q^{1,1}\,
R(y, \rho)\,,\label{Dinv}
\ee
where
\be
R(y, \rho) = \frac{1+ [(\rho-1)\,y - 1](y+1)^{\rho-1}}
{\rho\,(\rho-1)\, y^2}\,. \label{Dinv2}
\ee
For the ``canonical'' dimension $k=1\; (\rho = 2)$, we have
$R(y, \rho) = \frac 12$
that yields the free action. For any other non-zero and finite (``anomalous'')
value of $\rho$, we get
a non-trivial self-interaction. For instance, for $k=2\; (\rho =1)$
\be
R(y, 1) = \frac{1}{y} - \frac{\ln(1+y)}{y^2}\,. \label{WZW1}
\ee
This function, like \p{WZW}, is regular in $y$ in the vicinity of $y=0$.
A straightforward calculation (with doing the double harmonic integral
in its course) yields the following expression for the corresponding
metric function
$G_{scale}(f)$:
\be
G_{scale}(f) = \int du dv\, \frac{1- \hat{y}}{(1+\hat{y})^3} =
\frac{1}{\sqrt{\tilde{f}^2 - (\tilde{f}\cdot c)^2}}\,\arctan
\left[\frac{\sqrt{\tilde{f}^2 - (\tilde{f}\cdot c)^2}}
{1 + (\tilde{f}\cdot c)}\right]. \label{101}
\ee
We have chosen $c^2 =1$ in \p{101}, since the action \p{Dinv},
being scale-invariant,
does not depend on the norm of $c^{i\,a}$. The expression \p{101} manifests
the property
that the $SU(2)$ symmetries acting on the indices $i, a$ and, hence, the whole
super ${\bf h}(2)$ symmetry are broken in the action \p{Dinv}.
We have explicitly
checked that \p{101} (like \p{100}) obeys the 4-dimensional Laplace
equation \p{Lap}.

Note that for $k=0\; (\rho = \infty)$
no scale-invariant action can be constructed. Indeed,
in this case both $q^{1,1}$ and $\tilde{q}^{1,1}$
transform under $D$ as scalars of weight zero, so
there is no way to cancel the non-invariance
of the integration measure $\mu^{-2,-2}$.

Thus for the multiplet ${\bf (4,8,4)}$, irrespective of the  precise value
of the conformal dimension $k$
(or $\rho = 2/k$), there exists the unique  action \p{WZW}
which is invariant under the ${\cal N}{=}8$ super Heisenberg group,
but not under the $1D$ dilatations.
On the other hand, at any finite and non-vanishing $k$
one can construct superfield actions \p{Dinv}, \p{Dinv2}
which respect the scale invariance, but not invariance under the full
${\cal N}{=}8$ super Heisenberg group (they are  still invariant under
those two $SU(2)$ automorphism
groups with parameters in \p{param} which act on the underlined doublet indices).
The full conformal (and superconformal) invariance can be achieved
after coupling the rigid $q^{1,1}$ actions to the appropriate
${\cal N}{=}8, 1D$ supergravity, and it
emerges as a part of {\it local} ${\cal N}{=}8$ supersymmetry.

\subsection{Potential terms}
Let us also discuss the potential-type term of the superfield $q^{1,1}$
in the bi-harmonic superspace formalism
in the form
\be
S_{pot} \sim \,\int \mu^{-2,-2}\, \ti\, \ta\, C_{\und i\, \und a}\,
q^{1,1}\,. \label{POT}
\ee
Despite the presence of explicit $\theta$ s, this term is invariant
under Poincar\'e
${\cal N}{=}8$ supersymmetry, as a consequence of the defining constraints
\p{hc}.
Adding it (or its analog for few $q^{1,1}$) to the general sigma-model type
superfield action \p{s1gen} for $n=1$ (or for generic $n$) produces,
after eliminating
auxiliary fields, scalar potential which is fully specified by
the bosonic target
space metric (like in the ${\cal N}{=}(4,4), 2D$ case \cite{IS}).
Let us discuss
possible invariances of \p{POT}  under other transformations constituting
the supergroup
discussed in Subsect. 3.1.
\vspace{5mm}

{\it A) Dilatation transformations}.

On the Grassmann coordinates and superfield $q^{1,1}$
these transformations are realized as
\be
\delta \ti = \alpha_0\, \ti\,, \quad
\delta \ta = \alpha_0\, \ta\,, \quad
\delta q^{1,1} = k\, \alpha_0\, q^{1,1}\,, \quad
\delta \mu^{-2,-2} = -2 \alpha_0\, \mu^{-2,-2}\,.
\ee
Then the variation of \p{POT} takes the form
\be
\delta S_{pot} = k\, \alpha_0\, \,\int \mu^{-2,-2}\,
\ti\, \ta\,C_{\und i\, \und a}\, q^{1,1}\,.
\ee
The invariance of the potential term under these transformations
can be achieved only
for $k=0$ (or $\rho = \infty$). Since no invariant sigma-model type action
can be constructed in this particular case, we conclude that no scale-invariant
actions of the multiplet ${\bf (4, 8, 4)}$ with potential terms exist.

\vspace{5mm}

{\it B) $SU(2)$ transformations with parameters
$\lambda^{(ik)}$ and $\lambda^{(ab)}$}\,.\\

The transformation properties of the superfield $q^{1,1}$ and Grassmann
coordinates
under the action of these $SU(2)$ groups have the form
\bea
&&
\delta \ti = \ti\, \lambda^{(ij)} u^1_i\, u^{-1}_j\,, \quad
\delta \ta = \ta\, \lambda^{(ab)} v^1_a\, v^{-1}_b\,,\nn\\
&&
\delta q^{1,1} = \Big(\,\lambda^{(ij)} u^1_i\, u^{-1}_j
+ \lambda^{(ab)} v^1_a\, v^{-1}_b\, \Big)\, q^{1,1}\,.
\eea
The variation of the action has the form
\be
\delta S_{pot} =  \int \mu^{-2,-2}\, \Big (\,\lambda^{(ij)} u^1_i\, u^{-1}_j
+ \lambda^{(ab)} v^1_a\, v^{-1}_b\, \Big)\,
\ti\, \ta\, C_{\und i\, \und a}\,
q^{1,1}\,.
\ee
After integrating over $\theta$'s, one gets
\be
\delta S_{pot} = m\, \int dt\, du\, dv \Big (\,\lambda^{(ij)} u^+_i\, u^-_j
+ \lambda^{(ab)} v^+_a\, v^-_b\, \Big)\,
C_{\und i\, \und a}\, F^{\und i\, \und a}\,,
\ee
which is vanishing by harmonic integration.

\vspace{5mm}

{\it C)  $SU(2)$ transformations with parameters
$\lambda^{(\und i\, \und k)}$ and $\lambda^{(\und a\, \und b)}$}\,.\\

On the Grassmann coordinates and superfield $q^{1,1}$ these transformations
are realized as
\be
\delta \ti = \theta^{1,0\, \und k}\, \lambda^{\und i\,)}_{(\, \und k}\,, \quad
\delta \ta = \theta^{0,1\, \und b}\, \lambda^{\und a\,)}_{(\, \und b}\,, \quad
\delta q^{1,1} = 0\,.
\ee

The variation of the potential term  under these transformations
has the form
\be
\delta S_{pot} = \int \mu^{-2,-2}\,
\Big (\,\theta^{1,0\, \und k}\, \lambda^{\und i\,)}_{(\, \und k}
\theta^{0,1\, \und a}\,
+ \theta^{1,0\, \und i}\theta^{0,1\, \und b}\,
\lambda^{\und a\,)}_{(\, \und b}\,\Big )\,
C_{\und i\, \und a}\,q^{1,1}\,.
\ee
This variation cannot be made vanishing separately for both considered $SU(2)$.
However, for the choice of $C_{\und i\, \und a} = \eps_{\und i\, \und a}$
which
breaks the direct
product of these $SU(2)$ down to the diagonal $SU(2)$, the transformations
of the latter
(with the identification $\lambda^{(\,\und i}_{\;\und k\,)} =
\delta^{\und i}_{\und a}\, \delta^{\und b}_{\und k}\,
\lambda^{(\,\und a}_{\; \und b)}$)
leave \p{POT} invariant. Thus, the superfield potential term \p{POT}
can be made invariant under
three out of four automorphism $SU(2)$ symmetries. It can be shown
that this term
is not invariant
under the ``conformal'' supersymmetry with parameters
$\lambda^{i\, \und i}, \lambda^{a\, \und a}$
and, hence, under the central charge generator $Z$.

\setcounter{equation}{0}

\section{${\cal N}$=8, 1D  supergravity}

\subsection{${\cal N}$=8 SG from preserving bi-harmonic  analyticity}

By analogy with the ${\cal N}{=}(4,4)$, $2D$ case \cite{BI} we assume
that the fundamental group of ${\cal N}{=}8\;,$ $1D$ conformal SG is
represented by the following diffeomorphisms of the analytic
harmonic $SU(2)\times SU(2)$ superspace:
\bea
&&\delta \zeta^\mu = \Lambda^\mu (\zeta, u,v)\,,\;\;
\delta u^{1}_i = \Lambda^{2,0}(\zeta, u,v)\, u^{-1}_i\,,\;\;
\delta v^{1}_a = \Lambda^{0,2}(\zeta, u,v)\, v^{-1}_a\,,  \nonumber \\
&& \delta u^{-1}_i = \delta v^{-1}_a =0\,. \label{sgg}
\eea
Here $\zeta^\mu = (t, \theta^{1,0\,
\underline{k}}, \theta^{0,1\, \underline{b}})$
and the gauge parameters $\Lambda^\mu$, $\Lambda^{2,0}$,
$\Lambda^{0,2}$ are arbitrary functions
over the {\it whole} bi-harmonic analytic superspace
${\bf AR}^{(1+2+2|4)}$\,.

The
analyticity-preserving harmonic derivatives $D^{2,0}$ and $D^{0,2}$
defined in \p{harm.der}
are covariantized by introducing appropriate analytic vielbeins
\bea
D^{2,0} \Rightarrow  \nabla^{2,0} &=& D^{2,0} + H^{2,0\,\mu}
\partial_\mu
+ H^{4,0}\partial^{-2,0} +
H^{2,2}\partial^{0,-2} \nonumber \\
&\equiv & D^{2,0} +H^{2,0\,M}\partial_M \,,
\nonumber \\
D^{0,2} \Rightarrow  \nabla^{0,2} &=&
D^{0,2} + H^{0,2\,\mu}\partial_\mu
+ \tilde{H}^{2,2}\partial{}^{-2,0} + H^{0,4}\partial{}^{0,-2} \nonumber \\
&\equiv & D^{0,2} +H^{0,2\,M}\partial_M \,,
\label{covdII}
\eea
where we used the notation
\bea
&& M = (\mu, (2,0), (0,2)), \quad \partial_M = (\partial_\mu,
\partial{}^{-2,0},\partial{}^{0,-2}),\nonumber \\
&& \partial{}^{-2,0} = u^{-1\,i}
\frac{\partial}{\partial u^{1\,i}}\,, \quad
\partial{}^{0,-2} = v^{-1\,a}
\frac{\partial}{\partial v^{1\,a}}
\eea
and separated the flat parts of the vielbein components in
front of $\partial_{t}$ in $\nabla^{2,0}$ and $\nabla^{0,2}$.
In eqs. (\ref{covdII}) all vielbeins are
analytic ${\cal N}{=}8$\,, $1D$
superfields
$$
H^{2,0\,M}=H^{2,0\,M}(\zeta,u,v)\,,\;\;
H^{0,2\,M}=H^{0,2\,M}(\zeta,u,v)\,.
$$
The flat limit is achieved by putting
them equal to zero.
The $U(1)$ charge-counting operators $D^0_u$ and $D^0_v$ retain their
flat form (\ref{harm.der}).

Once again, in analogy with the consideration in ref. \cite{BI},
we postulate for
$\nabla^{2,0}$\,, $\nabla^{0,2}$ the following transformation law under the
${\cal N}{=}8$ SG group (\ref{sgg}):
\be \label{tranD}
\delta \nabla^{2,0} = -\Lambda^{2,0} D^0_u\,,\;\;\;
\delta \nabla^{0,2} = -\Lambda^{0,2} D^0_v\,,
\ee
whence
\bea
\delta H^{2,0} &=& \nabla^{2,0} \Lambda - 2i
\Lambda^{1,0\, \und i}\theta^{1,0}_{\und i}\,,
\nonumber \\
\delta H^{3,0\,\underline{i}} &=& \nabla^{2,0}
\Lambda^{1,0\,\underline{i}}
- \Lambda^{2,0}\theta^{1,0\;\underline{i}}\,, \;\;
\delta H^{2,1\,\underline{a}} \;=\;
\nabla^{2,0} \Lambda^{0,1\,\underline{a}}\,, \nonumber \\
\delta H^{4,0} &=& \nabla^{2,0}\Lambda^{2,0}\,,\;\;
\delta H^{2,2} \;=\; \nabla^{2,0}\Lambda^{0,2}\,,
\label{hdir20} \\
\delta H^{0,2} &=& \nabla^{0,2} \Lambda
-2i \Lambda^{0,1\, \und a}\theta^{0,1}_{\und a}\,, \nonumber \\
\delta H^{1,2\,\underline{i}} &=&
\nabla^{0,2} \Lambda^{1,0\,\underline{i}}\,,
\;\;
\delta H^{0,3\,\underline{a}} \;=\; \nabla^{0,2}
\Lambda^{0,1\,\underline{a}} - \Lambda^{0,2}
\theta^{0,1\,\underline{a}}\,,
\nonumber \\
\delta \tilde{H}^{2,2} &=& \nabla^{0,2}\Lambda^{2,0}\,,\;\;
\delta H^{0,4} \;=\; \nabla^{0,2}\Lambda^{0,2}\,.
\label{hdir02}
\eea

We wish to generalize the notion of the $\bf{(4,8,4)}$
analytic superfield $q^{1,1}$ to
the curved case. To this end, we need to find a correct generalization of the
defining constraints (\ref{hc}) and the ``superconformal'' transformation
laws (\ref{param}), \p{TRANq}, \p{LAMBD}.

In order to generalize the transformation laws of $q^{1,1}$ (\ref{TRANq}),
(\ref{LAMBD}) to the
curved case, we introduce two additional {\it independent} analytic gauge functions
$$
\Lambda_L(\zeta, u,v) = \lambda_L(t) + ... \,,\;\;
\Lambda_R(\zeta, u,v) = \lambda_R (t) + ...
$$
and ascribe the following transformation laws to $q^{1,1}$
\be \label{qutranloc}
\delta q^{1,1} = (\Lambda_L + \Lambda_R) q^{1,1}\,.
\ee
We can call these transformations the ``$U(1)$ weight'' or ``central charge''
ones, in order
to distinguish them from the trivial harmonic $U(1)$ phase transformations.
We normalize
the left and right central charges $J_L$ and $J_R$ so that
\be \label{quweight}
J_L q^{1,1} = J_R q^{1,1} = q^{1,1}\,.
\ee
At this stage, the $U(1)$ weight analytic parameters $\Lambda_L$\,, $\Lambda_R$
are entirely unrelated to those of coordinate transformations.

Such a relation naturally comes out, as a result of choosing
the appropriate transformation law for the $U(1)$ weight-covariantized
harmonic derivatives and fixing a proper gauge.

We covariantize $\nabla^{2,0}$\,, $\nabla^{0,2}$ by introducing
four analytic superfield $U(1)$ connections
${\cal H}^{2,0}_L(\zeta,u,v)$\,,
${\cal H}^{2,0}_R(\zeta,u,v)$\,, ${\cal H}^{0,2}_L(\zeta,u,v)$\,,
${\cal H}^{0,2}_R(\zeta,u,v)$
\bea
\nabla^{2,0} &\Rightarrow & {\cal D}^{2,0} =
\nabla^{2,0} +  {\cal H}^{2,0}_L J_L + {\cal H}^{2,0}_R J_R \nonumber \\
\nabla^{0,2} &\Rightarrow & {\cal D}^{0,2} =
\nabla^{0,2} +  {\cal H}^{0,2}_L J_L + {\cal H}^{0,2}_R J_R \,,
\label{u1covd}
\eea
and postulate the following transformation laws for ${\cal D}^{2,0}$\,,
${\cal D}^{0,2}$\,:
\bea
\delta {\cal D}^{2,0} &=& -\Lambda^{2,0}\, (D^0_u - J_L) -
\nabla^{2,0}\Lambda_L\,J_L - \nabla^{2,0}\Lambda_R\,J_R\,, \nonumber \\
\delta {\cal D}^{0,2} &=& -\Lambda^{0,2}\, (D^0_v - J_R)-
\nabla^{0,2}\Lambda_L\,J_L - \nabla^{0,2}\Lambda_R\,J_R \,.
\label{tranu1D}
\eea
The transformation laws of the vielbeins in $\nabla^{2,0}$\,,
$\nabla^{0,2}$ do not change,
while the newly introduced $U(1)$ connections are transformed as
\bea
\delta {\cal H}^{2,0}_L &=& \Lambda^{2,0} - \nabla^{2,0}\Lambda_L \,, \;\;
\delta {\cal H}^{2,0}_R \;=\; - \nabla^{2,0}\Lambda_R \,, \nonumber \\
\delta {\cal H}^{0,2}_L &=& - \nabla^{0,2}\Lambda_L \,,
\;\;\delta {\cal H}^{0,2}_R \;=\; \Lambda^{0,2} - \nabla^{0,2}\Lambda_R \,.
\label{tranu1con}
\eea
The ${\cal D}^{2,0}$
and ${\cal D}^{0,2}$ derivatives of the analytic
superfield  $\Phi^{p,q}$ with the left and right $U(1)$ weights
equal to $l$ and $r$ are transformed as follows:
\bea
\delta {\cal D}^{2,0}\, \Phi^{p,q} &=& -\Lambda^{2,0}\,(p-l)\,\Phi^{p,q} +
(l \Lambda_L + r \Lambda_R)\, {\cal D}^{2,0}\, \Phi^{p,q}\,, \nonumber \\
\delta {\cal D}^{0,2}\, \Phi^{p,q} &=& -\Lambda^{0,2}\,(q-r)\,
\Phi^{p,q} +
(l \Lambda_L + r \Lambda_R)\, {\cal D}^{0,2}\, \Phi^{p,q} \,.
\label{tranpqph}
\eea
We see that, only provided $p=l$\,, $q=r$\,, these derivatives are actually
covariant, i.e. they transform as the superfield $\Phi^{p,q}$ itself.
But this is precisely what happens for $q^{1,1}$\,, which possesses
$J_L = J_R =1$\,. Therefore,
as the appropriate curved generalization of the constraints (\ref{hc})
we choose the following ones:
\bea
{\cal D}^{2,0} q^{1,1} &=& (\nabla^{2,0} +
{\cal H}^{2,0}_L + {\cal H}^{2,0}_R)\, q^{1,1}
\;=\; 0 \,, \nonumber \\
{\cal D}^{0,2} q^{1,1} &=& (\nabla^{0,2} +
{\cal H}^{0,2}_L + {\cal H}^{0,2}_R)\, q^{1,1}
\;=\; 0 \,. \label{curqucons}
\eea

Note that the primary reason for the choice of transformation laws of
${\cal D}^{2,0}$\,, ${\cal D}^{2,0}$ in the form (\ref{tranu1D})
is the will to solder the coordinate transformations with the $U(1)$ weight
transformations, so as to eventually ensure a correct flat limit.
Indeed, from eqs. (\ref{tranu1con}) it follows
that the connections ${\cal H}^{2,0}_L$\,, ${\cal H}^{0,2}_R$ can be entirely
gauged away, thereby establishing the sought relation
\be  \label{basgauge}
{\cal H}^{2,0}_L = {\cal H}^{0,2}_R = 0 \;\;\Rightarrow \;\;
\Lambda^{2,0} = \nabla^{2,0}\Lambda_L\,,
\;\; \Lambda^{0,2} = \nabla^{0,2}\Lambda_R \,.
\ee
In what follows we will stick to this gauge.

An important consequence of the presence of two independent harmonic
constraints in the definition of the ${\cal N}{=}8$ superfield $q^{1,1}$\,,
eqs.
(\ref{curqucons}), is the integrability condition
\be  \label{intqu}
[{\cal D}^{2,0}, {\cal D}^{0,2}]\,q^{1,1} =0 \,.
\ee
It is easy to see that the direct generalization of the flat
condition $[D^{2,0}, D^{0,2}]=0$\,, namely
$$
[{\cal D}^{2,0}, {\cal D}^{0,2}]=0\,,
$$
is not covariant under (\ref{tranu1D}). The covariant version of this
constraint is as follows:
\be \label{constrD}
[{\cal D}^{2,0}, {\cal D}^{0,2}] =
- H^{2,2}(D^0_v - J_R) + \tilde{H}^{2,2}
(D_u^0 - J_L)\,.
\ee
It is evident that eq. (\ref{intqu}) is automatically satisfied, as
a consequence of (\ref{constrD}). This constraint implies
\bea
\tilde{H}^{2,2} = -\nabla^{2,0}\,{\cal H}^{0,2}_L\,, \;\;\;
H^{2,2} = - \nabla^{0,2}\,{\cal H}^{2,0}_R
\label{constrh22}
\eea
and
\be \label{consnab}
[\nabla^{2,0}, \nabla^{0,2}] = - H^{2,2} D^0_v +
\tilde{H}^{2,2} D^0_u\,.
\ee
From the latter relation one deduces the constraints on the analytic
vielbeins
\bea
\nabla^{2,0} H^{0,2} - \nabla^{0,2} H^{2,0} + 2i H^{2,1\, \und a}
\theta^{0,1}_{\und a} - 2i
H^{1,2\, \und i} \theta^{1,0}_{\und i} &=& 0 \,,\nonumber \\
\nabla^{2,0} H^{1,2\,\underline{i}} - \nabla^{0,2}
H^{3,0\,\underline{i}} - \tilde{H}^{2,2}
\theta^{1,0\,\underline{i}}&=& 0
\,, \nonumber \\
\nabla^{2,0} H^{0,3\,\underline{a}} - \nabla^{0,2}
H^{2,1\,\underline{a}}
+ H^{2,2}\theta^{0,1\,\underline{a}} &=& 0 \,,
\nonumber \\
\nabla^{2,0} H^{0,4} - \nabla^{0,2} H^{2,2} &=& 0\,, \nonumber \\
\nabla^{2,0} \tilde{H}^{2,2} - \nabla^{0,2} H^{4,0} &=& 0\,.
\label{constrH}
\eea

We do not know how to solve
(\ref{constrh22}), (\ref{constrH}) via unconstrained superfield
prepotentials. In order to single out the irreducible
field representation carried by vielbeins and $U(1)$ connections,
we keep to another strategy.
Namely, we use the initial gauge freedom, in order to gauge away from
these objects
as many components as possible, then substitute the resulting
expressions into the constraints
and solve the latter in this Wess-Zumino type gauge.
Eventually it turns out that
the solution exists, it is unique, and it is not reduced to a pure gauge.
The superfield constraints prove to be purely kinematic: they do
not imply any differential conditions or
equations of motion for the remaining fields, so the eventual gauge
field representations are off shell. The full nonlinear solution
of these general case constraints will be given elsewhere. Here, we limit
ourselves to the linearized level. This is quite sufficient
for revealing the irreducible field contents of the SG theory under
consideration. The full nonlinear solution will be presented below for
a simplified truncated version of ${\cal N}{=}8$ SG, corresponding to that
subgroup of \p{sgg} which doest not touch the harmonic variables.

In the present case one can choose the Wess-Zumino gauge in several different ways,
the basic criterion for one or another choice being the desire to
simplify the constraints (\ref{constrh22}), (\ref{constrH}) as much as
possible. As the starting step it is convenient to choose
the gauge (\ref{basgauge})
and the following additional ones:
\bea
H^{0,3\,\underline{a}} = 0\,, \;\;
H^{0,4} = 0\,. \label{gauge2}
\eea
These gauges restrict in certain way the original gauge parameters. At the
considered linearized level (\ref{gauge2}) give rise to
the following relations:
\bea
D^{0,2}\Lambda^{0,1\,\und a} - D^{0,2}\Lambda_R\, \theta^{0,1\,\und a} = 0\,, \;\;\;
(D^{0,2})^2 \Lambda_R = 0\,,\label{rel1}
\eea
which partly fix the $v$ dependence of the relevant gauge parameters.

There still remains a freedom
associated with the surviving harmonic dependence, and it can be used
to further gauge away some of
the field components in the double harmonic expansion of the remaining
vielbeins $H^{2,0}$\,, $H^{0,2}$\,, $H^{2,1\,\underline{a}}$\,,
$H^{1,2\,\underline{i}}$\,, $H^{3,0\,\underline{i}}$\,, $H^{4,0}$ and the
$U(1)$ connections ${\cal H}^{2,0}_R$\,, ${\cal H}^{0,2}_L$\,. This use of
the residual gauge freedom should be combined with solving the linearized
versions of the off-shell constraints \p{constrH}. After this
straightforward, though
somewhat cumbersome analysis (it is similar to that performed
in \cite{BI}) most of
the field components turn out to be eliminated, and one ends up with
the following final linearized form of the analytic vielbeins and gauge
superparameters:
\bea
H^{3,0\,\underline{i}} &=& H^{4,0} = 0\,, \nn \\
 H^{1,2\,\underline{i}} &=& i(\theta^{0,1})^2
\left[ h^{i\underline{i}}u^{1}_i + \theta^{1,0\,\underline{k}}
\left( h^{(\underline{i}}_{\;\;\underline{k})} +
\frac12 \delta^{\underline{i}}_{\underline{k}}\, \Gamma -
\delta^{\underline{i}}_{\underline{k}}\, l^{(ij)}u^{1}_i
u^{-1}_j \right) \right. \nn \\
&& \left. +\, (\theta^{(1,0)})^2\left( -i\partial_t{h}^{i\underline{i}}
+ \frac12 l^{i\underline{i}}\right)u^{-1}_i\right]\,, \nn \\
H^{2,1\,\underline{a}} &=& i(\theta^{1,0})^2\left[ h^{a\underline{a}}v^{1}_a
+ \theta^{0,1\,\underline{b}}
\left( h^{(\underline{a}}_{\;\;\underline{b})} +
\frac12 \delta^{\underline{a}}_{\underline{b}}\, \tilde{\Gamma} -
\delta^{\underline{a}}_{\underline{b}}\, r^{(ab)}v^{1}_a
v^{-1}_b \right) \right. \nn \\
&& \left.  +\, (\theta^{0,1})^2\left( -i\partial_t{h}^{a\underline{a}}
+ \frac12 r^{a\underline{a}}\right)v^{-1}_a\right], \nn \\
H^{2,0} &=& i(\theta^{1,0})^2\left[ h +
2i \theta^{0,1\,\underline{b}} h^a_{\underline{b}}v^{-1}_a -i(\theta^{0,1})^2
r^{(ab)}v^{-1}_av^{-1}_b \right] \,, \nn \\
H^{0,2} &=& i(\theta^{0,1})^2\left[ \tilde{h}
+ 2i \theta^{1,0\,\underline{i}} h^i_{\underline{i}}u^{-1}_i
-i(\theta^{1,0})^2
l^{(ij)}u^{-1}_iu^{-1}_j \right] \,, \nn \\
{\cal H}^{0,2}_L &=& i(\theta^{0,1})^2\left[ l + l^{(ik)}u^{1}_iu^{-1}_k
 + \theta^{1,0\,\underline{i}} l_{\underline{i}}^ku^{-1}_k
 -i (\theta^{1,0})^2\partial_t{l}^{(ik)}u^{-1}_iu^{-1}_k\right]\,, \nn \\
{\cal H}^{2,0}_R &=& i(\theta^{1,0})^2\left[ r + r^{(ab)}v^{1}_av^{-1}_b
 + \theta^{0,1\,\underline{a}} r_{\underline{a}}^b v^{-1}_b
 -i (\theta^{0,1})^2\partial_t{r}^{(ab)}v^{-1}_av^{-1}_b\right]\,,
\eea
\bea
\Lambda &=& \lambda + 2i \theta^{0,1\,\underline{a}}
\lambda^a_{\underline{a}} v^{-1}_a
+ 2i \theta^{1,0\,\underline{i}}\lambda^i_{\underline{i}} u^{-1}_i
+ i(\theta^{0,1})^2\omega^{(ab)}v^{-1}_av^{-1}_b \nn \\
&& +\, i(\theta^{1,0})^2\phi^{(ik)}u^{-1}_iu^{-1}_k\,, \nn \\
\Lambda^{1,0\,\underline{i}} &=& \lambda^{i\underline{i}} u^{1}_i
+ \theta^{1,0\,\underline{k}}\left[ \lambda^{(\underline{i}}_{\;\;
\underline{k})} +
\delta^{\underline{i}}_{\underline{k}} \left(\frac12 \gamma +
\phi^{(ik)}u^{1}_iu^{-1}_k\right)\right] \nn \\
&& -\, (\theta^{1,0})^2\left( i\partial_t{\lambda}^{i\underline{i}}
+ \frac12\phi^{i\underline{i}}\right)u^{-1}_i\,, \nn \\
\Lambda^{0,1\,\underline{a}} &=& \lambda^{a\underline{a}} v^{1}_a
+ \theta^{0,1\,\underline{b}}\left[ \lambda^{(\underline{a}}_
{\;\;\underline{b})} +
\delta^{\underline{a}}_{\underline{b}} \left(\frac12 \tilde\gamma
+ \omega^{(ab)}v^{1}_av^{-1}_b\right)\right] \nn \\
&& -\,(\theta^{0,1})^2\left(i\partial_t{\lambda}^{b\underline{a}}
+ \frac12\omega^{b\underline{a}}\right)v^{-1}_b, \nn \\
\Lambda_L &=& \phi + \phi^{(ik)}u^{1}_iu^{-1}_k + \theta^{1,0\,
\underline{i}} \phi^i_{\underline{i}} u^{-1}_i
-i (\theta^{1,0})^2\,\partial_t{\phi}^{(ik)}u^{-1}_iu^{-1}_k\,, \nn \\
\Lambda_R &=& \omega + \omega^{(ab)}v^{1}_av^{-1}_b
+ \theta^{0,1\,\underline{a}} \omega^a_{\underline{a}}
v^{-1}_a
-i (\theta^{0,1})^2\partial_t{\omega}^{(ab)}v^{-1}_av^{-1}_b\,.
\eea
All fields and component gauge parameters in these formulas are functions only of $t$.
The fields $h\,, \tilde h\,, \Gamma\,, \tilde\Gamma$ are still related
by the equation following
from the first constraint in \p{constrH}. In the linearized form it reads
\be
\partial_t(\tilde h - h) + (\tilde\Gamma - \Gamma) = 0\,. \label{EQlin}
\ee

The residual linearized transformation laws
\be
\delta h = \partial_t{\lambda} - \gamma \,, \quad
\delta\tilde{h} = \partial_t{\lambda}
- \tilde\gamma\,, \quad
\delta\Gamma = \partial_t{\gamma}\,, \quad
\delta\tilde\Gamma = \partial_t{\tilde\gamma}
\ee
show that the ``would-be'' einbeins $h$ and $\tilde{h}$ can be
fully gauged away by two local parameters $\gamma$ and $\tilde\gamma$
\be
h = \tilde h = 0 \quad \Rightarrow \quad \gamma = \tilde\gamma
= \partial_t{\lambda}\,.
\ee
In this gauge eq. \p{EQlin} implies
\be
\Gamma = \tilde\Gamma\,, \quad \delta\Gamma = \partial^2_t{\lambda}\,.
\label{AffConn}
\ee
Thus, the interesting peculiarity of this ${\cal N}{=}8\,, 1D$
``Weyl multiplet'' is that
it contains no einbein in the maximally restricted Wess-Zumino gauge.
Instead, it
contains a sort of $1D$ affine connection with the transformation
law \p{AffConn}.
This is, of course, a consequence of the original presence of two independent
local scale parameters $\gamma$ and $\tilde\gamma$ in the considered
${\cal N}{=}8\,, 1D$
``conformal SG'' group and two corresponding compensating $1D$ fields $h$
and $\tilde h$ in the analytic vielbeins.
In the $2D$ case \cite{BI} the analogous compensating fields are
collected among
the original 4 components of the zweibein. After gauging them away,
one is still left
with the ``conformal zweibein'' having two components that are
shifted by derivatives
of the two $d=2$
diffeomorphisms parameters. In $1D$ case nothing remains from
the ``conformal einbein''.
We shall see below how the standard $1D$ einbein, the transformation
of which starts
with a shift by $\partial_t{\lambda}$\,,
reappears within our approach.

For what follows, it is instructive to write down the minimal gauge
fields representation,
which we have arrived at, and the corresponding gauge parameters
(the numbers within round brackets
to the right stand for the ``engineering'' dimension in mass units):
\bea
\underline{\mbox{bosons}}: &&\Gamma(t)\,(1),\;
h^{(\underline{i}\underline{k})}(t)\, (1),\;
h^{(\underline{a}\underline{b})}(t)\, (1), \; l^{(ik)}(t)\, (1), \;
r^{(ab)}(t) \,(1),
\;l(t)\,(1), \; r(t)\,(1), \nn \\
&&\partial_t{\lambda}(t)\,(0), \;
\lambda^{(\underline{i}\underline{k})}(t)\, (0),\;
\lambda^{(\underline{a}\underline{b})}(t)\, (0),\; \phi^{(ik)}(t)\, (0),\;
\omega^{(ab)}(t)\, (0), \;
\phi(t)\, (0), \; \omega(t)\,(0), \nn \\
\underline{\mbox{fermions}}: &&h^{i\underline{i}}(t)\, (1/2), \;\;\;
h^{a\underline{a}}(t)\, (1/2), \;\;\;
l^{i\underline{i}}(t)\, (3/2), \;\;\; r^{a\underline{a}}(t)\, (3/2), \nn \\
&&\lambda^{i\underline{i}}(t)\,(-1/2),\;\lambda^{a\underline{a}}(t)\,
(-1/2),\; \phi^{i\underline{i}}(t)\, (1/2), \;
\omega^{a\underline{a}}(t)\, (1/2)\,. \label{Table1}
\eea
The linearized transformation of any gauge field starts with
``$\partial_t$''  of the corresponding gauge
parameter.

Thus, in the considered Wess-Zumino gauge we are left with  15 bosonic gauge fields
(one connection, two central charge (or ``$U(1)$ weight'') gauge fields
and twelve gauge fields for
four mutually commuting $SU(2)$ groups) and 16 fermionic gauge fields
(eight Poincar\'e gravitini and eight ``conformal''
gravitini). At first sight, this mismatch between bosonic
and fermionic degrees of freedom seems to be
a signal of inconsistency. However, the number of gauge parameters
is the same, which means that locally
we deal with the $(0 + 0)$ off-shell representation. This is also a
characteristic feature of $2D$
Weyl supermultiplets \cite{Schout,BI}. Later on, we shall argue that
the formal equality of fermionic
and bosonic fields can be restored after coupling this gauge field
representation to the appropriate
$1D$ compensating superfields.

Actually, in order to be able to construct manifestly invariant superfield
couplings of ${\cal N}{=}8$ SG  multiplets to ${\cal N}{=}8$ matter we
need one more ingredient, namely, an analytic density which would transform
so as to cancel the transformation of the analytic
superspace integration measure
$\mu^{-2, -2}$\,. Indeed, the full local
group \p{sgg} does not leave $\mu^{-2,-2}$ invariant
\be \label{sgmeas}
\delta \mu^{-2,-2} = \left[(-1)^{P(\rho)}\partial_\rho\Lambda^\rho +
\partial^{-2,0}\Lambda^{2,0} + \partial^{0,-2}
\Lambda^{0,2}\right]\mu^{-2,-2}
\equiv \tilde{\Lambda}\; \mu^{-2,-2}\,,
\ee
where $P(\rho)$ is 0 for bosonic and 1 for fermionic indices.

Defining the objects
\be  \label{Gamma}
\Gamma^{2,0} = (-1)^{P(M)} \partial_M H^{2,0\;M}\,,\;\;\;
\Gamma^{0,2} = (-1)^{P(M)} \partial_M H^{0,2\;M}\,,
\ee
one finds them to transform as
\begin{equation}
\delta \Gamma^{2,0} = \nabla^{2,0} \tilde{\Lambda}\,,\;\;\;
\delta \Gamma^{0,2} = \nabla^{0,2} \tilde{\Lambda}\,.
\end{equation}
and to satisfy, as a consequence of the constraints (\ref{constrH}), the
condition
\be  \label{constrgam}
\nabla^{2,0} \Gamma^{0,2} - \nabla^{0,2} \Gamma^{2,0} = 0\,.
\ee
It is easy to show that (\ref{constrgam}) implies
\be  \label{gamsigm}
\Gamma^{2,0} = \nabla^{2,0} \Sigma (\zeta,u,v)\,,\;\;
\Gamma^{0,2} = \nabla^{0,2} \Sigma (\zeta,u,v)\,.
\ee
Once again, with making use of the constraints (\ref{constrH}),
$\Sigma (\zeta, u,v)$ can be
expressed in terms of the original SG multiplet (up to an unessential
additive constant) and shown to transform as
\be \label{trsigma}
\delta \Sigma = \tilde{\Lambda}\,.
\ee
Hence the quantity
\be \label{compvol}
\Omega  \equiv e^{-\Sigma} \,,\;\;\;\delta \Omega  = -\tilde{\Lambda} \,
\Omega
\ee
is the sought object compensating the non-invariance
of the measure. Due to the property
\be \label{propgam}
(\,\nabla^{2,0} + \Gamma^{2,0}\,)\, \Omega = 0\,,\;\;\;
(\,\nabla^{0,2} + \Gamma^{0,2}\,)\, \Omega = 0
\ee
one can still integrate by parts with respect
to the {\it covariantized} harmonic derivatives. Indeed, for any
analytic function $F(\zeta, u, v)$ the integral
$$
\int \mu^{-2,-2}\, \Omega\, \nabla^{2,0}F(\zeta,u,v)\,,
$$
up to total {\it ordinary} derivatives with respect to the analytic superspace coordinates
(including the harmonic ones), reduces to
$$
- \int \mu^{-2,-2}\, (\nabla^{2,0} + \Gamma^{2,0})\,\Omega\, F(\zeta,u,v)
= 0
$$
(the same is true for $\nabla^{0,2}$)\,.

To close this subsection, we wish to point out that the
${\cal N}{=}8$ SG supergroup can be treated
as gauging of the maximal finite-dimensional subgroup which preserves
the flat form of the harmonic derivatives (see Subsect. 3.1). Conversely,
this subgroup can be recovered as that subgroup of \p{sgg}, \p{qutranloc} which survives after
putting the analytic supervielbeins and ``central charge'' connections equal to zero.
The general SG group \p{sgg} (as well as its truncation \p{sgg1}
considered below) contains arbitrary
reparametrizations of $t_A$ constituting the Virasoro algebra in which the finite-dimensional
$1D$ conformal algebra  $so(1,2)$ forms a subalgebra. It is plausible, that at least some of
the ${\cal N}{=}8\,, 1D$ superconformal groups listed in \cite{VPr} also form subalgebras in
\p{sgg} (or \p{sgg1}). For the time being, it is not clear to us what
is possible significance of such
finite-dimensional superconformal subalgebras in the context of the bi-harmonic formulation
of ${\cal N}{=}8\,, 1D$ SG considered here.

\subsection{Simplified ${\cal N}$=8 SG}
Here we consider, at the full nonlinear level, a truncated version of
the ${\cal N}{=}8$ SG of the previous section. This simplified version
amounts to the following choice of local supergroup:
\bea
\delta \zeta^\mu = \Lambda^\mu (\zeta, u,v)\,,\;\;
\delta u^{\pm 1}_i = \delta v^{\pm 1}_a = 0\,. \label{sgg1}
\eea

The corresponding covariantized harmonic derivatives, transformation rules and
constraints can be obtained by setting
\be
H^{4,0} = H^{0,4} = H^{2,2} = \tilde{H}^{2,2} = 0 \label{Trun0}
\ee
in the appropriate formulas of the previous Subsection:
\be
\nabla^{2,0} = D^{2,0} + H^{2,0\,\mu} \partial_\mu\,,
\quad \nabla^{0,2} =
D^{0,2} + H^{0,2\,\mu}\partial_\mu\,,  \label{Trun1}
\ee
\bea
\delta H^{2,0} &=& \nabla^{2,0} \Lambda - 2i
\Lambda^{1,0\,\und i}\theta^{1,0}_{\und i}\,,\;\;
\delta H^{0,2} = \nabla^{0,2} \Lambda
-2i \Lambda^{0,1\, \und a}\theta^{0,1}_{\und a}\,, \nn \\
\delta H^{3,0\,\underline{i}} &=& \nabla^{2,0}
\Lambda^{1,0\,\underline{i}}\,, \;\;
\delta H^{0,3\,\underline{a}} = \nabla^{0,2}
\Lambda^{0,1\,\underline{a}}\,, \nn \\
\delta H^{2,1\,\underline{a}} &=&
\nabla^{2,0} \Lambda^{0,1\,\underline{a}}\,, \;\;
\delta H^{1,2\,\underline{i}} =
\nabla^{0,2} \Lambda^{1,0\,\underline{i}}\,, \label{Trun2}
\eea
\bea
\nabla^{2,0} H^{0,2} - \nabla^{0,2} H^{2,0} +
2i H^{2,1\,\und a}\theta^{0,1}_{\und a}-
2i H^{1,2\, \und i} \theta^{1,0}_{\und i} &=& 0 \,,\nonumber \\
\nabla^{2,0} H^{1,2\,\underline{i}} - \nabla^{0,2}
H^{3,0\,\underline{i}} &=& 0
\,, \nonumber \\
\nabla^{2,0} H^{0,3\,\underline{a}} - \nabla^{0,2}
H^{2,1\,\underline{a}}
 &=& 0 \,. \label{Trunconstr}
\eea

In order to find the full solution of \p{Trunconstr}, we shall proceed as
in the previous Subsection while deriving the linearized solution. We firstly
impose the appropriate gauge. This time it will be convenient to choose
it as
\bea
&& H^{0,3\,\underline{a}} = 0 \;\; \Rightarrow \;\;
\nabla^{0,2}\Lambda^{0,1\,\underline{a}} = 0 \nn \\
&& H^{0,2} = 0 \;\;\Rightarrow \;\;
\nabla^{0,2}\Lambda - 2i \Lambda^{0,1\, \und a}\theta^{0,1}_{\und a} = 0\,.
\eea
Using the gauge parameter $\Lambda^{1,0\,\underline{i}}$\,,
one can also reduce $H^{1,2\,\underline{i}}$ to
the form
\be
H^{1,2\,\underline{i}} = i(\theta^{0,1})^2 h^{1,0\,\underline{i}}
(t,u, \theta^{1,0\,\und i})\,,
\ee
after which $\nabla^{0, 2}$ is radically shortened
\be
\nabla^{0, 2} = D^{0,2} +
i(\theta^{0,1})^2 h^{1,0\,\underline{i}}\partial_{1,0\,\underline{i}}\,.
\ee
Next, one substitutes the $\theta$-expansion of all remaining vielbein
components into \p{Trunconstr}
and combines the resulting set of equations for fields given on the manifold $(t,u,v)$ with
the remaining gauge freedom,
in order to eliminate as many component fields as possible and to fully fix
the harmonic dependence of the
surviving ones. This analysis is essentially simplified, due to the fact that,
in fixing the Wess-Zumino gauges,
we still can use only the shifting parts of the gauge transformations,
i.e. the linearized form
of the latter. The resulting irreducible form of the vielbeins which solves
\p{Trunconstr} reads
\bea
H^{2,1\,\underline{a}} &=& i (\theta^{1,0})^2
\left\{ h^{a\underline{a}}v^{1}_a
+ \theta^{0,1\,\underline{b}}
\left[h_{(\underline{b}}^{\;\;\underline{a})} +
\frac{1}{2}\, \delta^{\underline{a}}_{\underline{b}}\left(\Gamma
-2i\, h^i_{\underline{k}}\,h^{l\underline{k}}\,u^{1}_iu^{-1}_l\right)\right]
\right. \nn \\
&&  -\,i(\theta^{0,1})^2\left[ \partial_t{h}^{a\underline{a}}
+ \left(\Gamma -
2i\, h^i_{\underline{k}}\,h^{l\underline{k}}\,u^{1}_iu^{-1}_l\right)
h^{a\underline{a}}\right]
v^{-1}_a \bigg \} \nn \\
&& -i \theta^{1,0 \,\underline{i}}\theta^{0,1\,\underline{a}}\,
h^i_{\underline{i}}\, u^{1}_i +
2 (\theta^{0,1})^2\theta^{1,0 \,\underline{i}}\,
h^i_{\underline{i}}\,h^{a\underline{a}}\, u^{1}_i v^{-1}_a\,,
\nn \\
H^{3,0\,\underline{i}} &=& 2 (\theta^{1,0})^2\theta^{0,1\,\underline{a}}\,
h^{i\underline{i}}\,h^{a}_{\underline{a}}\,
u^{1}_iv^{-1}_a\,, \quad H^{2,0} =
-2(\theta^{1,0})^2\theta^{0,1\,\underline{a}}\,
h^{a}_{\underline{a}}\,v^{-1}_a\,, \nn \\
h^{1,0\,\underline{i}} &=& h^{i\underline{i}}\,u^{1}_i
+\theta^{1,0 \,\underline{k}}
\left[h_{(\underline{k}}^{\;\;\underline{i})} +
\frac{1}{2}\, \delta^{\underline{i}}_{\underline{k}}\left(\Gamma
-2i\, h^i_{\underline{k}}\,h^{t\underline{k}}\,
u^{1}_iu^{-1}_t\right)\right] \nn \\
&& - i(\theta^{1,0})^2\left[\partial_t{h}^{i\underline{i}}
+ \frac{1}{2}\, \Gamma h^{i\underline{i}}
- h_{\;\;\underline{k})}^{(\underline{i}}\, h^{i\underline{k}}
+ \frac{i}{3}\left(h^i_{\underline{k}}\,h^{k\underline{k}}\right)
h_k^{\underline{i}}\right] u^{-1}_i\,.
\label{FullH}
\eea
The field $\Gamma$ and all fields $h$ here are unconstrained
functions of $t$\,. The resulting
irreducible field content can be summarized as the following truncation
of the table \p{Table1}:
\bea
\underline{\mbox{bosons}}: &&\Gamma(t)\,(1),\;
h^{(\underline{i}\underline{k})}(t)\, (1),\;
h^{(\underline{a}\underline{b})}(t)\, (1),  \nn \\
&&\partial_t{\lambda}(t)\,(0), \;  \lambda^{(\underline{i}\underline{k})}(t)\, (0),\;
\lambda^{(\underline{a}\underline{b})}(t)\, (0), \nn \\
\underline{\mbox{fermions}}: &&h^{i\underline{i}}(t)\, (1/2), \;\;\;
h^{a\underline{a}}(t)\, (1/2),  \nn \\
&&\lambda^{i\underline{i}}(t)\,(-1/2),\;\lambda^{a\underline{a}}(t)\, (-1/2)\,. \label{Table2}
\eea
As in the previous ``master'' case, we are facing the $(0 + 0)$ off-shell gauge multiplet at this step.

We also need the precise form of the
relevant superconnections $\Gamma^{2,0}$ and $\Gamma^{0,2}$
defined by \p{Gamma} (taking into account
the truncation \p{Trun0})
\bea
\Gamma^{2,0} &=& -2i \theta^{1,0 \,\underline{i}}h^{i}_{\underline{i}}\,
u^{1}_i
-i(\theta^{1,0})^2\left(\Gamma - 2i h^i_{\underline{k}}\,h^{l\underline{k}}\,
u^{1}_iu^{-1}_l\right)
\left(1 + 2i\theta^{0,1\,\underline{b}}h^a_{\underline{b}}\,
v^{-1}_a\right)\,, \nn \\
\Gamma^{0,2} &=& -i(\theta^{0,1})^2\left(\Gamma - 2i
h^i_{\underline{k}}\,h^{l\underline{k}}\,u^{1}_iu^{-1}_l \right)
\nn \\
&& +2 (\theta^{0,1})^2 \theta^{1,0 \,\underline{i}}
\left[\partial_t{h}^{i}_{\underline{i}}
+ \frac{1}{2}\, \Gamma h^{i}_{\underline{i}}
- h_{(\underline{i}\underline{l})}\, h^{i\underline{l}}
+ \frac{i}{3}\left(h^i_{\underline{k}}\,h^{k\underline{k}}\right)
h_{k\underline{i}}\right] u^{-1}_i\,.
\label{Gammy}
\eea
It is easy to explicitly check that these $\Gamma^{2,0}$ and $\Gamma^{0,2}$
satisfy the integrability
condition
$$
\nabla^{2,0}\Gamma^{0,2} -\nabla^{0,2}\Gamma^{2,0} = 0\,,
$$
and to show that
$$
\Gamma^{2,0} =\nabla^{2,0}\Sigma\,, \quad \Gamma^{0,2} = \nabla^{0,2}\Sigma \,,
$$
with
\be
\Sigma = \ln e - 2i \theta^{1,0\,\underline{i}}\,h^{i}_{\underline{i}}\,
u^{-1}_i
- (\theta^{1,0})^2 h^i_{\underline{k}}\,h^{k\underline{k}}\,
u^{-1}_iu^{-1}_k\,,
\quad \Gamma = - e^{-1}\partial_t{e}\,.
\label{Omega}
\ee
The new object $e(t)$ can be identified with the standard einbein.

\subsection{Multiplet (4,8,4) in ${\cal N}$=8 SG background}
Our aim is to construct a locally supersymmetric extension of the free superfield action for
the $(\bf{4, 8, 4})$ multiplet $q^{1,1}$\,.
We assume that under the truncated
${\cal N}{=}8$ SG group \p{sgg1}
this superfield transforms as
\be
\delta q^{1,1}(\zeta, u, v) = q^{1,1}{}'(\zeta',u,v) - q^{1,1}(\zeta,u,v)
= -\frac{1}{2}\, \tilde{\Lambda} q^{1,1}(\zeta,u,v)\,, \label{TrunTranq}
\ee
where
\be
\tilde{\Lambda} = \partial_t{\Lambda} -\partial_{1,0\,\underline{i}}\,
\Lambda^{1,0\,\underline{i}}
-\partial_{0,1\,\underline{a}}\,\Lambda^{0,1\,\underline{a}}\,,
\quad \delta \mu^{-2,-2}
= \tilde{\Lambda}\, \mu^{-2,-2}\,.
\ee
Then, the invariant action has the same form as in the flat case
\be
S = \int \mu^{-2,-2}\, q^{1,1} q^{1,1} \,. \label{ACT}
\ee
The crucial difference from the flat case is encoded in the locally
supersymmetric constraints
which should be now imposed on $q^{1,1}$
\be
\left(\nabla^{2,0} + \frac{1}{2}\, \Gamma^{2,0}\right) q^{1,1} = 0\,, \quad
\left(\nabla^{0,2} + \frac{1}{2}\, \Gamma^{0,2}\right) q^{1,1} = 0\,.
\label{Trunqconstr}
\ee
They are manifestly covariant under the ${\cal N}{=}8$ SG transformations
\p{sgg}, \p{Trun2} and \p{TrunTranq}.

In order to find the component action, one should rewrite \p{Trunqconstr}
as a set of harmonic equations for
the component fields in the $\theta$ expansion of $q^{1,1}$
\bea
q^{1,1} &=& f^{1,1} + \theta^{1,0\,\underline{i}}\psi_{\underline{i}}^{0,1} +
\theta^{0,1\,\underline{a}}\psi_{\underline{a}}^{1,0}
+(\theta^{1,0})^2 M^{-1,1}
+ (\theta^{0,1})^2 N^{1,-1}\nn \\
&& + \theta^{1,0\,\underline{i}}\theta^{0,1\,\underline{a}}
F_{\underline{i}\,\underline{a}} +
(\theta^{0,1})^2\theta^{1,0\,\underline{i}}\xi_{\underline{i}}^{0,-1} +
(\theta^{1,0})^2\theta^{0,1\,\underline{a}}\xi_{\underline{a}}^{-1,0} +
(\theta^{0,1})^2(\theta^{1,0})^2 d^{-1,-1}\,, \label{Expanq}
\eea
where all fields ``live'' on the manifold $(t,u^{\pm 1}_i, v^{\pm 1}_a)$\,.
These
component harmonic equations can easily be derived from \p{Trunqconstr} using
the explicit expressions \p{FullH}, \p{Gammy}. We explicitly present only the
few simplest ones
\bea
&& (\mbox{a}) \;\;\partial^{2,0}f^{1,1} = \partial^{0,2}f^{1,1} = 0 \;\Rightarrow \;
f^{1,1} = f^{ia}(t)u^{1}_iv^{1}_a\,, \nn \\
&& (\mbox{b}) \;\;\partial^{2,0} \psi^{0,1}_{\underline{i}}
-i h^i_{\underline{i}} u^{1}_i f^{ka}u^{1}_kv^{1}_a = 0\,, \;\;
\partial^{0,2}\psi^{0,1}_{\underline{i}} = 0 \; \Rightarrow \nn \\
&& \psi^{0,1}_{\underline{i}} =
\psi^{a}_{\underline{i}}(t)v^{1}_a  +
i h^i_{\underline{i}}(t) f^{ka}(t) u^{1}_{(i}u^{-1}_{k)}v^{1}_a\,, \nn \\
&& (\mbox{c}) \;\;\partial^{2,0} \psi^{1,0}_{\underline{a}} =
\partial^{0,2} \psi^{1,0}_{\underline{a}} = 0
\; \Rightarrow \;\psi^{1,0}_{\underline{a}} = \psi^{i}_{\underline{a}}(t)u^{1}_i\,, \nn \\
&& (\mbox{d})\;\;\partial^{2,0} F^{\underline{i}\,\underline{a}} =
\partial^{0,2} F^{\underline{i}\,\underline{a}} = 0 \; \Rightarrow \;
F^{\underline{i}\,\underline{a}} = F^{\underline{i}\,\underline{a}}(t)\,.
\eea
The remaining equations look more complicated, but can be straightforwardly solved
as well. Things
are somewhat simplified, recalling that our aim is to compute the component action
which follows from \p{ACT}
\be
S = \int dt\, dv\, du\, \left( 2 f^{1,1}d^{-1,-1} -
\psi^{0,1\,\underline{i}}\xi^{0,-1}_{\underline{i}} -
\psi^{1,0\,\underline{a}}\xi^{-1,0}_{\underline{a}} + 2 M^{-1,1}N^{1,-1} -
\frac{1}{4}\, F^{\underline{i}\,\underline{a}}F_{\underline{i}\,\underline{a}}
\right). \label{ACT2}
\ee
We can integrate in \p{ACT2} by parts with respect to harmonic derivatives, e.g. representing
$f^{1,1}= \partial^{2,0}f^{-1,1}\,, \; \psi^{1,0}_{\underline{a}} =
\partial^{2,0}\psi^{-1,0}_{\underline{a}}$, etc. As a result, one can directly use in
\p{ACT2} the harmonic equations for the higher components in \p{Expanq}, without explicitly
solving these equations. After some work, we obtain the final rather simple
answer for the component
Lagrangian in \p{ACT2} for the case of coupling to the considered version of ${\cal N}{=}8$ SG
\be
S = \int dt\, \left({\cal D}f^{ia}\,{\cal D}f_{ia}
+ \frac{i}{2}\psi^{i\underline{a}}\,
\nabla\psi_{i\underline{a}}
+ \frac{i}{2}\tilde\psi^{\underline{i}a}\,\nabla\tilde\psi_{\underline{i}a} -
\frac{1}{4}\, F^{\underline{i}\,\underline{a}}
F_{\underline{i}\,\underline{a}} \right) \label{ACT3}
\ee
where
\bea
{\cal D}f^{ia} &=& \partial_t{f}{}^{ia} - \frac{1}{2}\,\Gamma\,f^{ia} +
h^{i\underline{k}}\,\tilde{\psi}^a_{\underline{k}}
+ h^{a\underline{b}}\,\psi^i_{\underline{b}}\,, \nn \\
\nabla\psi_{i\underline{a}} &=& \partial_t{\psi}_{i\underline{a}} +
\tilde{h}_{(\underline{a}}^{\;\;\underline{b})}\,\psi_{i\underline{b}}\,,\;\;
\nabla\tilde{\psi}_{\underline{i}a} = \partial_t{\tilde{\psi}}_
{\underline{i}a} +
\tilde{h}_{(\underline{i}}^{\;\;\underline{k})}\,
\tilde{\psi}_{\underline{k}a}\,,
\eea
and
\be
\tilde{\psi}{}^{\underline{i}a}  = \psi^{\underline{i}a} +
\frac{i}{2}\,h^{\underline{i}k}f_k^a\,, \;\;
\tilde{h}_{(\underline{i}}^{\;\;\underline{k})}
= h_{(\underline{i}}^{\;\;\underline{k})} +
ih^i_{\underline{i}}\,h_i^{\underline{k}}\,,\;\;
\tilde{h}_{(\underline{a}}^{\;\;\underline{b})} =
h_{(\underline{a}}^{\;\;\underline{b})} +
2i h^a_{\underline{a}}\,h_a^{\underline{b}}\,.
\ee

It is straightforward to find the full residual gauge transformations which preserve
the Wess-Zumino gauge \p{FullH} and leave invariant the action \p{ACT3}.
For simplicity, we shall present only
that part of the local supersymmetry transformations of the component fields
of $q^{1,1}$ which
is of zeroth order in the SG fields
\bea
&&\delta f^{ia} = -\lambda^{i\underline{i}}\,\tilde{\psi}^a_{\underline{i}} -
\lambda^{a\underline{a}}\,\psi^i_{\underline{a}}\,, \;
\delta\tilde{\psi}^a_{\underline{i}} = 2i\lambda^{i}_{\underline{i}}\,
\partial_t{f}^a_i -
\lambda^{a\underline{a}}F_{\underline{i}\,\underline{a}}\,, \;
\delta\psi^i_{\underline{a}} = 2i\lambda^{a}_{\underline{a}}\,
\partial_t{f}^i_a +
\lambda^{i\underline{i}}F_{\underline{i}\,\underline{a}}\,, \nn \\
&&\delta F_{\underline{i}\,\underline{a}} = -2i\lambda^{i}_{\underline{i}}\,
\partial_t{\psi}_{i\underline{a}}
+ 2i\lambda^{a}_{\underline{a}}\,\partial_t{\tilde{\psi}}_{\underline{i}a}\,.
\eea
It is easy to check that the response of the pure matter part
of \p{ACT3} against
these transformations is exactly cancelled by the shift part of
the gravitini transformations ($\delta h^{i\underline{i}} =
\partial_t{\lambda}^{i\underline{i}}$\,,
$\delta h^{a\underline{a}}=\partial_t{\lambda}^{a\underline{a}}$)
in the part which is linear
in $h^{i\underline{i}}$\,, $h^{a\underline{a}}$
\be
2 \partial_t{f}^{ia}\,\tilde{\psi}^{\underline{i}}_a \,h_{i\underline{i}} +
2 \partial_t{f}^{ia}\,\psi^{\underline{a}}_i\, h_{a\underline{a}}\,.
\ee

Summarizing, the component action \p{ACT3} enjoys local ${\cal N}{=}8$
supersymmetry (gauge fields
$h^{i\underline{i}}\,, h^{a\underline{a}}$)\,, $1D$ diffeomorphisms
(gauge field $\Gamma$)
and two local $SU(2)$ symmetries realized on the underlined
doublet indices (gauge fields
$\tilde{h}^{(\underline{a}\underline{b})}\,,
\tilde{h}^{(\underline{i}\underline{k})}$)\,.
Besides, it respects two global $SU(2)$ symmetries realized
on the non-underlined
doublet indices. Thus the specificity of this version of
${\cal N}=8$ SG is that
local $SU(2)$ groups are realized only on fermions and auxiliary fields of the multiplet
${\bf (4,8,4)}$ (like in the version of ${\cal N}{=}4\,, 1D$ SG treated
in \cite{{Sor1},{Sor2}}).
Respectively, the corresponding gauge fields couple
to currents which involve only fermions. However, as follows
from the results of \cite{IS1},
there should exist other versions of the same ${\bf(4,8,4)}$ multiplet,
in which physical
bosonic fields have different $SU(2)$ assignments and carry at least one underlined
doublet index. Hence, in the corresponding locally ${\cal N}{=}8$
supersymmetric actions the $SU(2)$
gauge fields should non-trivially couple to physical bosonic fields as well. We plan to study
such models elsewhere.

As a last topic, we discuss two mechanisms of gaining the standard einbein
in this approach.

One of them consists in identifying the einbein with the norm of the
$4$-vector $f^{ia}$\,, taking
into account that $f^{ia}$ undergoes local weight transformations
with the parameter
$\partial_t{\lambda}$\,. In order to turn on this mechanism, one should
vary \p{ACT3} with respect to $\Gamma$ as
an auxiliary field and express $\Gamma$ in terms of other fields from the corresponding
algebraic equation of motion
\bea
\delta \Gamma: \quad {\cal D}f^{ia}f_{ia} = 0 \;\;\Rightarrow \;\;
\Gamma &=& -e^{-1}\partial_t{e}  +
2\,e^{1/2}\left(h^{i\underline{k}}\,\tilde{\psi}^a_{\underline{k}}\,
\hat{f}_{ia} +
h^{a\underline{b}}\,\psi^i_{\underline{b}}\,\hat{f}_{ia}\right) \nn \\
&\equiv & -e^{-1}\partial_t{e} + 2\,\Omega^{ia}\hat{f}_{ia}\,, \label{GammaEq}
\eea
where we split $f^{ia}$ into the radial and angular parts as
\be
f^{ia} = e^{-1/2}\hat{f}^{ia}\,, \quad \hat{f}^{ia}\hat{f}_{ia} = 1\;
\Rightarrow \; f^2 = e^{-1}\,.
\ee
After substituting this back into \p{ACT3}, one obtains
\be
S = \int dt\, \left\{\,\frac{1}{e}\left[\partial_t{\hat{f}} -
(\Omega \cdot \hat{f})\hat{f} + \Omega \right]^2
+\frac{i}{2}\psi^{i\underline{a}}\,\nabla\psi_{i\underline{a}}
+ \frac{i}{2}\tilde\psi^{\underline{i}a}\,\nabla\tilde\psi_{\underline{i}a} -
\frac{1}{4}\, F^{\underline{i}\,\underline{a}}
F_{\underline{i}\,\underline{a}}\,\right\}\,. \label{ACT4}
\ee
Actually, one can repeat the same procedure off shell, redefining $\Gamma$ as
\be
\Gamma =-e^{-1}\partial_t{e} + 2\,\Omega^{ka}\hat{f}_{ka} + \tilde{\Gamma}\,.
\ee
After substituting this back into \p{ACT3}, one obtains just \p{ACT4} with the
addition
\be
 \int dt\,\frac{1}{e}\,\tilde{\Gamma}^2\,,\label{Add}
\ee
i.e. the on-shell expression \p{GammaEq} for $\Gamma$ simply amounts to the elimination of the
auxiliary field $\tilde{\Gamma}$
\be
\tilde{\Gamma} = 0\,.
\ee
The physical boson part of the sum \p{ACT4} + \p{Add}  is nothing but
the world-line covariant
action of the particle moving on the sphere $S^3 \sim SO(4)/SO(3)$
parametrized by $\hat{f}^{ia}$\,.

The resulting off-shell representation is a collection of gauge and ``matter'' fields.
The gauge field $(0+0)$ representation contains 7 bosonic gauge fields $e\,,
\tilde{h}^{(\underline{i}\underline{k})}\,,
\tilde{h}^{(\underline{a}\underline{b})}$ and eight
fermionic gauge fields $h^{i\underline{k}}\,, h^{a\underline{b}}$\,.
The ``matter'' sector is represented
by the set ${\bf (3, 8, 5)}$\,, with 3 physical $S^3$ bosonic fields
$\hat{f}^{ia}$\,, eight fermionic fields
$\psi^{i\underline{a}}\,, \tilde{\psi}^{\underline{i}a}$ and five auxiliary
fields $\tilde{\Gamma}\,,
F^{\underline{i}\,\underline{a}}$\,.
Hopefully, the latter set can be equivalently
understood as a nonlinear
version of the off-shell ${\cal N}{=}8\,, 1D$ multiplet
${\bf (3,8,5)}$ \cite{BIKL2}, an analog
of the corresponding nonlinear ${\cal N}{=}4, 1D$ multiplet
${\bf (3, 4,1)}$ \cite{IL,IKL2}.

An alternative mechanism of generating the einbein is through
the ``prepotential'' $\Sigma$
defined in \p{Omega}. In this case the off-shell einbein
is introduced ``by hands'', so the
set of matter fields remains intact, i.e. ${\bf( 4, 8, 4)}$\,. The gauge fields
representation is the same as in the previous case.
In this case we can redefine
$q^{1,1} = e^{-1/2 \Sigma}\, {q}^{1,1}_0$ where ${q}^{1,1}_0$ is a scalar
of zero weight and satisfies the simplified version of the constraints
\p{Trunqconstr} containing
no superfield connections $\Gamma^{2,0}\,, \Gamma^{0,2}$\,. In this approach
it is straightforward to construct a locally ${\cal N}{=}8$
supersymmetric version of the general sigma model
action \p{s1gen} for the ${\bf (4,8,4)}$ multiplets.
One simply should make the replacement
$\mu^{-2, -2} \;\Rightarrow \; \mu^{-2,-2}\, e^{- \Sigma}$ in \p{s1gen}
and take into account that
the involved superfields ${q}^{1,1\,M}_0$ satisfy the covariantized
constraints just mentioned.

Finally, we argue that the gauge representation \p{Table1} corresponding to the ``master''
${\cal N}{=}8$ SG group \p{sgg} can be reduced to the set
consisting of ${\bf (1, 8, 7)}$ fields,
upon an appropriate choice of the matter compensating superfield.

Let us assume that, as such, one can choose the ``extreme'' off-shell multiplet
${\bf (8,8,0)}$ \cite{GR} with ${\bf 7}$ bosons of dimension 0 and ${\bf 8}$ fermions
of dimension $1/2$
as compensating fields. We assume that one of the original ${\bf 8}$ bosonic field
is going to become the
einbein, as in the previous example, while the remaining ${\bf 7}$ fields
compensate one out of two
local central charge transformations and two out of four local $SU(2)$ symmetries.
The fermions have
the correct dimension to be compensators for ${\bf 8}$ local ``conformal''
supersymmetries (with parameters
$\phi^{i\underline{i}}(t)$ and $\omega^{a\underline{a}}(t)$).
What would remain in the gauge where all compensating fields
are put equal to zero is ${\bf 8}$
bosonic gauge fields (the einbein,
one gauge field for the remaining central charge and six gauge fields
for two remaining $SU(2)$) and
${\bf 8}$  Poincar\'e gravitini. The ``matter'' multiplet will comprise
${\bf 8}$ bosonic fields of dimension 1
($\tilde\Gamma = \Gamma + \partial_t e +...$, one former central charge gauge field
and ${\bf 6}$ former gauge fields for two entirely compensated
$SU(2)$ symmetries) and ${\bf 8}$ fermionic fields of dimension 3/2
(the former ``conformal'' gravitini).
So, in this case the full set of gauge fields could be organized
into the off-shell ${\bf (1,8,7)}$
multiplet (although the actual number of off-shell degrees of freedom
will still remain $(0 + 0)$, in view
of the full matching in the number of gauge fields and gauge parameters).
In order to check this conjecture,
one needs to find full nonlinear solution for the corresponding
supervielbeins and to construct
the invariant coupling of the ${\bf (8, 8, 0)}$ multiplet to this generic
${\cal N}{=}8$ SG background.
This is a good problem for a future consideration.
Another interesting problem is to establish a correspondence with
the ${\cal N}{=}8$ SG superfield formalism in the ordinary ${\cal N}{=}8\,,
1D$ superspace, as a reduction of the
analogous ${\cal N}{=}4$ constructions in $2D$ \cite{Gates1,Gates2}.

\section{Concluding remarks}
We finish with a brief summary of the paper and outlining some
further directions of the study.

One of our purposes was to show that the bi-harmonic analytic ${\cal N}{=}8$ superspace
provides an adequate
framework for describing ${\cal N}{=}8$ mechanics associated with
the off-shell multiplet ${\bf (4,8,4)}$\,.
We constructed the general superfield and component actions both for single
such multiplet and
for the case when a few multiplets are involved, and presented the relevant superfield
potential terms. We also identified the maximal superalgebra constituted by
those coordinate transformations
which do not affect the flat form of two commuting harmonic derivatives
preserving the bi-harmonic $1D$
Grassmann analyticity. This supergroup turned out to be a ${\cal N}{=}8$
superextension of the two-dimensional Heisenberg algebra {\bf h}(2)\,,
rather than any kind of ${\cal N}{=}8\,, 1D$
superconformal algebra. Such a superextension was not known before.
We constructed the corresponding
unique invariant action of one ${\bf(4,8,4)}$ multiplet and showed
that it is not scale-invariant. On the other hand,
there exists a one-parameter family of scale-invariant actions with
a non-trivial self-interaction which
do not respect invariance under the full ${\cal N}{=}8$ Heisenberg supergroup.

Another incentive of this paper was to formulate a non-propagating
${\cal N}{=}8\,, 1D$ supergravity in
the analytic bi-harmonic superspace, proceeding from the universal principle
of preserving harmonic Grassmann
analyticity and following the same tokens as in the ${\cal N}{=}(4,4)\,, 2D$ case \cite{BI}.
The most
general (``master'') version of such a theory was considered at the linearized level,
while for its simplified version,
with the harmonic variables being intact under the corresponding superdiffeomorphisms,
we found the full nonlinear
solution of the relevant constraints in WZ gauge. For this latter case
we presented the first example of
off-shell locally ${\cal N}{=}8$ supersymmetric action
of the multiplet ${\bf (4, 8,4)}$ and discussed
some peculiarity of such a system related to two ways of generating
the einbein gauge field which is not present
among the original gauge fields of the considered versions of ${\cal N}{=}8$ SG.

These results can be extended in several directions. An interesting,
though basically technical problem
is to establish the full nonlinear structure of the master version of
${\cal N}{=}8$ SG and to construct
the corresponding locally supersymmetric actions for the multiplet ${\bf (4,8,4)}$ in this SG
background. More ambitious project is to find out possible physical implications
of such actions, including the one presented in this paper. The $1D$ SG fields
are non-propagating
and serve as Lagrange multipliers for the
Hamiltonian constraints which after quantization become equations
of motion for spinning fields
on the space of bosonic moduli of the ``matter'' $1D$ multiplets (see \cite{{Sor1},{Sor2}}
and refs. therein). Following this line, the bosonic fields
of the ${\bf (4,8,4)}$ multiplet could
hopefully be treated as parameters of some compact manifold ${\cal K}^{4n}$
the dimension of which is multiple of 4
and which could arise in some compactification schemes in
higher-dimensional supergravities or string
theory. Then the ${\cal N}{=}8$ SG - ${\bf (4,8,4)}$ actions could be treated as describing a
``relativistic'' particle moving on the factor ${\cal K}^{4n}$
in the product of this ``internal'' manifold
by some non-compact manifold which represents the ``space-time'' part of the relevant
compactification, e.g. some AdS$_m$ manifold. For describing the motion of particle
on the full product
manifold, one clearly needs to add couplings of the world-line ${\cal N}{=}8$ SG
to some other ``matter''
${\cal N}{=}8\,, 1D$ multiplets the physical bosons of which would represent
the coordinates of
the non-compact factor just mentioned. The quantization of such an extended system
could produce
some interesting type of dynamical equations for fields with spin on the product manifold
as a background space-time. By analogy with the ${\cal N}{=}4$ case where the
multiplets ${\bf (1,4,3)}$ \cite{IKLev} were utilized
under similar circumstances \cite{Sor1,Sor2},
the minimal possibility in the ${\cal N}{=}8$ case is to incorporate the
off-shell ${\bf (1,8,7)}$ multiplets \cite{ABC} for the above purpose.\footnote{We
thank J. Buchbinder for suggesting such a possibility and discussions of these
and related issues.} Then an
interesting problem for the future study is to describe this multiplet in ${\cal N}{=}8$
bi-harmonic superspace and to construct its coupling to the versions of ${\cal N}{=}8$ SG
considered here. We hope that this description can be achieved using the method similar to the
one employed in \cite{IS1} for description of non-equivalent twisted multiplets in the
same analytic ${\cal N}{=}(4,4)\,, 2D$ HSS.

An independent interesting task is to explore possible dynamical models on the ${\cal N}{=}8$
Heisenberg supergroup and its various cosets,
in particular, to utilize the latter as target superspaces
for the appropriate superparticle models.
Due to the presence of the central charge $Z$, both in the bosonic and fermionic sectors,
such models are expected to admit one-dimensional Wess-Zumino terms and,
by the same reasoning as
in refs. \cite{{Japan},{IMT}}, to provide ${\cal N}{=}8$ superextension of two-dimensional
Landau problem. Such superextensions are known to bear a tight relation to non-commutative
``fuzzy'' supermanifolds, like the standard bosonic Landau problem is related to
non-commutative plane and two-dimensional ``fuzzy'' sphere.

\section*{Acknowledgements}

We thank J. Buchbinder, S.J. Gates, Jr. and S. Krivonos for interest
in the work and useful discussions.
This research was partially supported by the European Community's
Marie Curie Research Training Network
under contract MRTN-CT-2004-005104 Forces Universe, and by the grant
INTAS-00-00254. E.I. and A.S.
also acknowledge a support from RFBR grant, project  No 03-02-17440, and a grant of the
Heisenberg-Landau program. They thank the Directorate of LNF-INFN for
the kind hospitality at different
stages of this study.

\end{document}